\documentclass[twocolumn]{aastex631}

\shorttitle{Residual Abundances of MW Satellites}
\shortauthors{Hasselquist et al.}

\begin{document}

\title{2-process Model and Residual Abundance Analysis of the Milky Way Massive Satellites}

\author[0000-0001-5388-0994]{Sten Hasselquist}
\affiliation{Space Telescope Science Institute, 3700 San Martin Drive, Baltimore, MD 21218, USA}

\author[0000-0003-2969-2445]{Christian R. Hayes}
\affiliation{Herzberg Astronomy and Astrophysics Research Centre: Victoria, British Columbia, CA}
\affiliation{Space Telescope Science Institute, 3700 San Martin Drive, Baltimore, MD 21218, USA}

\author[0000-0001-9345-9977]{Emily J. Griffith}
\affiliation{Center for Astrophysics and Space Astronomy, Department of Astrophysical and Planetary Sciences, University of Colorado, 389 UCB, Boulder, CO 80309-0389, USA}

\author[0000-0001-7775-7261]{David Weinberg}
\affiliation{The Department of Astronomy and Center of Cosmology and AstroParticle Physics, The Ohio State University, Columbus, OH 43210, USA}

\author[0000-0001-8208-9755]{Tawny Sit}
\affiliation{Department of Astronomy, The Ohio State University, Columbus, OH 43210, USA}

\author[0000-0002-1691-8217]{Rachael L. Beaton}
\affiliation{Space Telescope Science Institute, 3700 San Martin Drive, Baltimore, MD 21218, USA}
\affiliation{Department of Physics and Astronomy, Johns Hopkins University, Baltimore, MD 21218, USA}

\author[0000-0003-1856-2151]{Danny Horta}
\affiliation{Center for Computational Astrophysics, Flatiron Institute, 162 5th Ave., New York, NY 10010, USA\\}

\begin{abstract}

The ``2-process Model'' is a promising technique for interpreting stellar chemical abundance data from large-scale surveys (e.g., SDSS-IV/V, GALAH), enabling more quantitative empirical studies of differences in chemical enrichment history between galaxies without relying on detailed yield and evolution models. In this work, we fit 2-process model parameters to (1) a luminous giant Milky Way (MW) sample and (2) stars comprising the Sagittarius Dwarf Galaxy (Sgr). We then use these two sets of model parameters to predict the abundances of 14 elements of stars belonging to the MW and in five of its massive satellite galaxies, analyzing the residuals between the predicted and observed abundances. We find that the model fit to (1) results in large residuals (0.1-0.3 dex) for most metallicity-dependent elements in the metal-rich ([Mg/H] $>$ -0.8) stars of the satellite galaxies. However, the model fit to (2) results in small or no residuals for all elements across all satellite galaxies. Therefore, despite the wide variation in [X/Mg]-[Mg/H] abundance patterns of the satellite galaxies, the 2-process framework provides an accurate characterization of their abundance patterns across many elements, but these multi-element patterns are systematically different between the dwarf galaxy satellites and the MW disks. We consider a variety of scenarios for the origin of this difference, highlighting the possibility that a large inflow of pristine gas to the MW disk diluted the metallicity of star-forming gas without changing abundance ratios. 

\end{abstract}


\section{Introduction} \label{sec:intro}

Understanding the detailed chemical abundance patterns for individual stars in a galaxy is essential for understanding their detailed histories of star formation, accretion, and outflows, for the subset of galaxies where such measurements can be obtained. The nucleosynthetic sites and timescales for element production are known for many chemical elements, and thus the details of the star formation history of a galaxy can be inferred from its detailed chemical abundance patterns. For example, the abundance of the $\alpha$ elements (O, Mg, Si, S, and Ca) as measured relative to the abundance of a heavier element, such as Fe (often expressed as [$\alpha$/Fe]), can provide some estimate of the star formation efficiency of a galaxy, with elevated or increasing abundance ratios indicating times of relatively vigorous star formation and deficient or decreasing abundance ratios signaling times of less vigorous star formation, with delayed Type Ia supernovae (SNe Ia) becoming more prominent sources of chemical enrichment \citep[e.g.,][]{Tinsley1979,Shetrone2003,Edvardsson1993,Hendricks2014,Nidever2020,Hasselquist2021}. The abundances of the elements produced in large quantities by the $s$-process in asymptotic giant branch (AGB) stars can indicate the relative contribution of these stars to the chemical evolution of a galaxy or globular cluster \citep[e.g.,][]{Tinsley1979,Kobayashi2006,Kobayashi2020}. Measuring abundance ratios of the elements produced in hydrostatic burning to those produced in explosive synthesis can probe some details of the initial mass function (IMF) of a galaxy at the time at which it formed those stars \citep[e.g.,][]{McWilliam2013,Hasselquist2017,Carlin2018,Ji2020}. Further detailed abundance patterns of even more elements can reveal aspects of stochastic sampling of the IMF \citep[e.g.,][]{Koch2008,Ji2020,Griffith2023} and/or assist in our understanding of the exact nature of less well-understood enrichment sources, such as the sources of the $r$-process \citep[e.g.,][]{Sneden2008,Ji2016,Lian2023}.

Despite numerous studies that have advanced our knowledge of how specific details of the star formation history (SFH) vary with galaxy mass and environment, powerful quantitative interpretations remain elusive. Ideally, one could use a parameterized chemical evolution model to fit multi-element abundances, each of which is sensitive to different aspects of a star formation history. Such efforts, while important and insightful, generally find that not every chemical element is able to be accurately predicted in the MW \citep[see e.g., ][]{Andrews2017,Rybizki2017,Kobayashi2020}. The failures are likely some combination of the abundance measurements themselves (e.g., not including corrections for non-local thermodynamic equilibrium, or ``NLTE''), incorrect or incomplete yield calculations, and unknown or exotic sources that are not explicitly accounted for in the models \citep[e.g., sub-Chandrasekhar mass Type Ia SNe;][]{Woosley1994,McWilliam2018}. These tools are further complicated by the fact that each study that measures abundances is typically affected by unconstrained systematic uncertainties, complicating how these abundance results are combined and compared.

One tool with potential to better quantify and understand differences in the detailed star formation history of galaxies is the ``2-process Model'' that was first described in \citet{Weinberg2019} and applied to chemical abundance measurements from hundreds of thousands of Milky Way (MW) stars observed by the Apache Point Observatory Galaxy Evolution Experiment \citep[APOGEE;][]{Majewski2017}. The major assumption of this model is that every chemical abundance in a star can be described as the sum of contributions to this element from a ``prompt'' process (typically assigned to Type II or ``core-collapse supernovae'', CCSNe) and a ``delayed'' process (typically assigned to SNe Ia). They divided the MW into two samples based on their [Fe/Mg]-[Mg/H] abundance patterns: the low-Ia (high-$\alpha$) sequence, which is a sequence comprised primarily or entirely by the ``prompt'' process, and the high-Ia (low-$\alpha$) sample, which is a sequence that formed from contributions from both processes. From the observed trends of [X/Mg] vs. [Mg/H] within these two populations, they inferred model parameters describing the relative contribution of each process to the observed elements, X. \citet{Weinberg2019} then used this model to predict the abundances of the other elements measured by APOGEE, confirming that the abundance patterns of these sequences for virtually all of these elements are identical throughout the Galactic disk \citep[see also e.g.,][]{Nidever2014,Hayden2015}, with follow-up works confirming that these chemical similarities extend to the Galactic Bulge and Inner Galaxy \citep[e.g.,][]{Zasowski2019,Griffith2021}. \citet{Griffith2022} found that the 2-process model is able to do similarly well in predicting the abundances of MW stars that have abundances derived from the Galactic Archaeology with HERMES survey \citep[GALAH;][]{Sheinis2015} data release 3 \citep[DR3][]{Buder2021} rather than APOGEE, obtaining similar results for elements common to both surveys and extending the 2-process model to elements not measured by APOGEE.

It is worth emphasizing that the 2-process model is \emph{not expected} to be a perfect description of high-dimensional abundance patterns; if anything, its accuracy in fitting these patterns in the MW disk and bulge is a surprise. The fact that the detailed abundance patterns of nearly all MW stars is well-fit by the 2-process model implies that the nucleosynthetic processes creating the elements and dispersing them throughout the ISM of the MW were largely invariant with time and Galactic location. In other words, the IMF-averaged yields of each process appear to result in the same chemical abundances at fixed metallicity ([Mg/H]) and relative contribution from each process (as traced by [Fe/Mg]). However, the small element-by-element residuals from a 2-parameter description are statistically significant \citep{Ting2022,Weinberg2022,Ratcliffe2023} and some stars observed by APOGEE show much larger residuals. Specifically, \citet{Weinberg2022} showed that stars belonging to the LMC and $\omega$~Cen exhibit large residuals between the observed and model-predicted abundances for many elements, a result confirmed and extended to other MW satellite galaxies by \citet{Sit2024} with a much larger sample after taking into account potential systematic abundance measurement errors. The magnitude, detailed form, and environmental dependence of these residuals from the 2-process model can provide empirical clues about enrichment processes and their impact in different galaxies and Galactic components.

Analyzing and interpreting these empirical clues remains a difficult task as it is still an open question as to whether these 2-process model abundance residuals indicate differences in the element yields of the prompt or delayed process in different systems, the role of a third or fourth process that has different relative amplitude in these systems, or a more fundamental breakdown of the assumptions that lead to the 2-process description in the first place. For example, we might expect elements primarily produced in AGB stars to show large 2-process model residuals for dwarf galaxies that are known have had significant contribution to their chemical enrichment from AGB stars, as this process is not separately accounted for in the 2-process model. However, for the elements in the MW that appear to be accurately predicted via a combination of CCSNe and Type Ia SNe, large 2-process model residuals in extra-Galactic systems taken at face value suggest actual differences in the IMF-averaged yields of one or both types of SNe. Of course, such differences need not necessarily imply differences in stellar astrophysics, as the residuals could be revealing a flaw in the model assumptions that are more obvious in systems with different galactic evolution. Therefore, a more in-depth study of what these residuals actually mean is warranted. 

Moreover, it is still unknown if the dwarf galaxy residual abundance patterns found in \citet{Sit2024} are shared across all dwarf galaxies, or if each galaxy exhibits its own unique residual abundance pattern. Stated more generally, are the residual abundance patterns of the dwarf galaxies indicative of a fundamental difference in how a MW-sized galaxy is formed vs. a dwarf galaxy, or do the residual abundance patterns of each depend uniquely on mass and environment? \citet{Hasselquist2021} attempted to answer this question without making use of the 2-process model, finding that, while the $\alpha$-element abundance patterns implied vast differences in the star formation rate as a function of time between these galaxies, the detailed abundance patterns of the other elements were much more difficult to interpret. For many elements observed in these galaxies, it was inconclusive if the variation in abundance patterns were signals of fundamental differences between the chemical enrichment processes that occurred, or, like in the MW, the abundance patterns are what are expected given the ratio of a prompt process to a delayed process, as traced by the [Mg/H] and [Fe/Mg] abundances of each galaxy.

In this work, we carefully select a MW sample that matches the stellar parameters of the satellite galaxy sample of \citet{Hasselquist2021}, and derive two separate sets of abundance vectors for the 2-process model, creating a ``MW 2-process model'' and a ``DG (dwarf galaxy) 2-process model''. We then use both models to predict the chemical abundances of the satellite galaxies, linking residuals to potential real differences in the detailed chemical evolution of these galaxies, issues with the 2-process model assumptions, or both. The data used and target selection are described in \autoref{sec:data}. The 2-process model and how we fit the model parameters are described in \autoref{sec:two_proc}. The model parameters that we derive are discussed in \autoref{sec:results_vectors}. Residual abundance results are presented in \autoref{sec:resid_abund_results} and discussed in \autoref{sec:disc}.

\section{Data and Target Selection} \label{sec:data}

We use observations from the Apache Point Observatory Galactic Evolution Experiment \citep[APOGEE][]{Majewski2017}, which was a near-infrared spectroscopic survey of hundreds of thousands of stars across the Milky Way as part of the third and fourth iterations of the Sloan Digital Sky Survey (SDSS-III: \citealt{Eisenstein2011} and SDSS-IV: \citealt{Blanton2017}). The APOGEE survey used two nearly-identical spectrographs \citep{Wilson2019}, one on the SDSS 2.5 meter telescope at Apache Point Observatory \citep{Gunn2006} and another on the 2.5 meter du Pont telescope at Las Campanas Observatory \citep{Bowen73}.

Throughout the lifetime of these surveys, the APOGEE spectrographs obtained high-resolution (R $\sim$ 22,000) $H$-band spectra of nearly 750,000 stars distributed throughout the Milky Way (MW) and its nearby satellite galaxies \citep[targeting descriptions are given in: ][]{Zasowski2013,Zasowski2017,Beaton2021,Santana2021}, from which stellar parameters and detailed chemical abundances were obtained using the APOGEE Stellar Parameters and Chemical Abundance Pipeline (ASPCAP, \citealt{Garcia-Perez2016}). ASPCAP uses the FERRE code (\citealt{AllendePrieto2006}) to match the normalized, RV-corrected, and sky-subtracted spectra \citep{Nidever2015} to a library of synthetic stellar spectra (\citealt{Zamora2015} with updates described in \citealt{Jonsson2020}), allowing for the determination of up to $\sim$ 20 chemical abundances. In this work, we use data from Data Release 17 \citep{dr172022}, which includes NLTE calculations for Na, Mg, K, and Ca \citep{Osorio2020}, along with Ce abundances from \citet{Cunha2017}. 

In this work, we analyze a sample of dwarf galaxies studied in \citet{Hasselquist2021}, which include the LMC, SMC, Sgr, Fnx, and the large fraction of the MW halo that is believed to have come from a single progenitor galaxy (e.g., \citealt{Nissen2010,Hayes2018a,Belokurov2018,Deason2018,Helmi2018}), which we refer to in this work as the \emph{Gaia} Sausage/Enceladus galaxy (GSE).  The selection of stars for each of these systems is described in detail by \citet{Hasselquist2021} and was based on \emph{Gaia} DR2 proper motions \citep{GaiaDR2}, APOGEE radial velocities, distances and orbital parameters from the \texttt{astroNN} code \citep{Leung2019}, and, in the case of the MCs, additional photometric selections were made to exclude obvious AGB stars and massive super giant stars.

We also select a MW sample that is used to derive the model parameters for the 2-process model. Because of small APOGEE systematic abundance trends with effective temperature and surface gravity \citep[see e.g.,][]{Jonsson2020,Griffith2021,Weinberg2022,Sit2024}, we are careful to select a MW sample that covers the same T$_{\rm eff}$  and $\log{g}$ distributions as the stars in the massive satellite sample. To select our MW sample, we first explicitly exclude the satellite galaxy members identified in \citet{Hasselquist2021}, and then start with the following selection criteria:
\begin{itemize}
    \item $[$Fe/H$]$ $<$ 0.2, as no dwarf galaxy has stars more metal-rich than this value,
    \item log(g) $<$ 3.0, as this removes obvious dwarf stars because main sequence stars are too faint for APOGEE to observe at the distances of these satellite galaxies,
    \item median S/N per pixel $>$ 80, and
    \item targeting flag \texttt{EXTRATARG} = 0 to ensure sample does not include any ``special targets''\footnote{see \url{https://www.sdss4.org/dr17/irspec/targets/}}, e.g., stars that were targeted based on their known or suspected membership in globular clusters (GCs). To be complete as possible, we also remove any stars that have been identified by \citet{Schiavon2024} to be in GCs that may have been serendipitously targeted by APOGEE. 
\end{itemize}

\begin{figure*}
    \includegraphics[width=\textwidth]{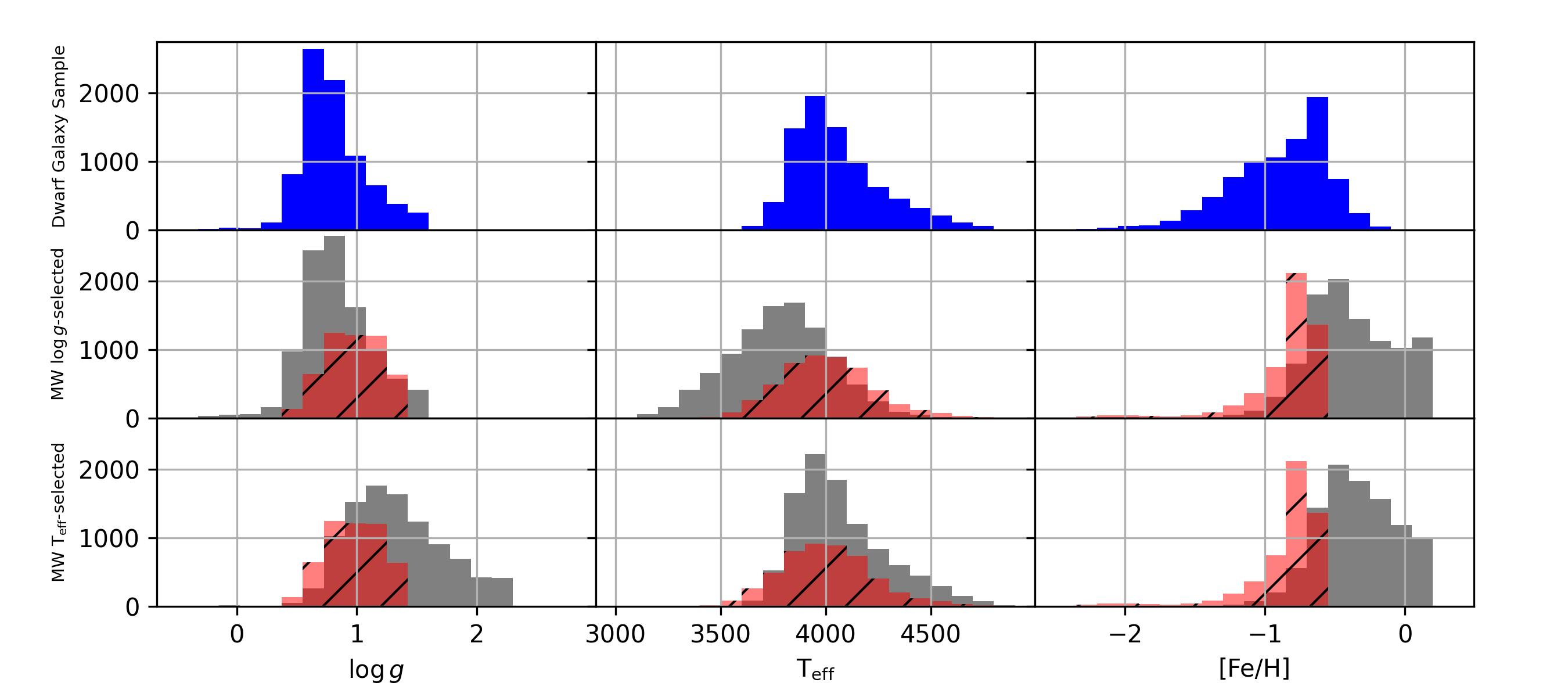}
    \caption{Stellar parameter distributions for the dwarf galaxy sample (top row) and MW selection sub-samples (bottom two rows). The middle row shows the stellar parameters of the MW sample obtained by selecting stars such that the $\log{g}$ distribution is the same as the dwarf galaxies (sub-sample 1). The bottom row is the same as the second row, but matching on T$_{\rm eff}$ (sub-sample 2). The MW ``metal-poor'' sample, as described in the text, is shown in the hatched red distribution.\label{fig:param_select}}
\end{figure*}

From this MW sample, which is comprised of $\sim$30,000 stars, we then create two separate MW sub-samples by randomly sub-selecting 10,000 MW stars that either match the T$_{\rm eff}$ distribution (sub-sample 1) or match the $\log{g}$ distribution (sub-sample 2) to that of the satellite galaxies. The distributions of stellar parameters for each of these selections, and how they compare to those of the satellite galaxy sample, are shown in  \autoref{fig:param_select}. In general, the satellite galaxy distributions are matched fairly well, but in the case of the $\log{g}$ distribution, there are not quite enough MW stars at each $\log{g}$ to return an exact distribution. Moreover, the metallicity distribution of the MW T$_{\rm eff}$-matched sample is slightly more metal-poor than the $\log{g}$-matched distribution. Therefore, we focus most of the results and analysis on the T$_{\rm eff}$-matched sample, showing some of the results from the $\log{g}$-matched sample in \autoref{sec:appendix}. Our main results and conclusions are not sample-dependent.

Because the APOGEE survey contains a relatively small fraction of stars at [Fe/H] $<$ -0.65, we supplement the above 10,000 star MW sample with \emph{all} stars observed by APOGEE with [Fe/H] $<$ -0.65 that meet the above criteria with an additional criterion of $\log{g}$ $<$ 1.6. Of course, this results in a sample that is largely contaminated from GSE stars that were not included in the \citet{Hasselquist2021} sample, as their sample was focused on purity rather than completeness. Therefore, we mitigate GSE contamination in this metal-poor sample by enforcing the following additional cuts on these metal-poor stars:
\begin{itemize}
    \item $[$Fe/H$]$ $<$ -0.65 \& [Mg/Fe] $>$ 0.05
    \item $[$Fe/H$]$ $<$ -0.95 \& [Mg/Fe] $>$ 0.27
\end{itemize}

\begin{figure*}
    \epsscale{1.2}
    \plotone{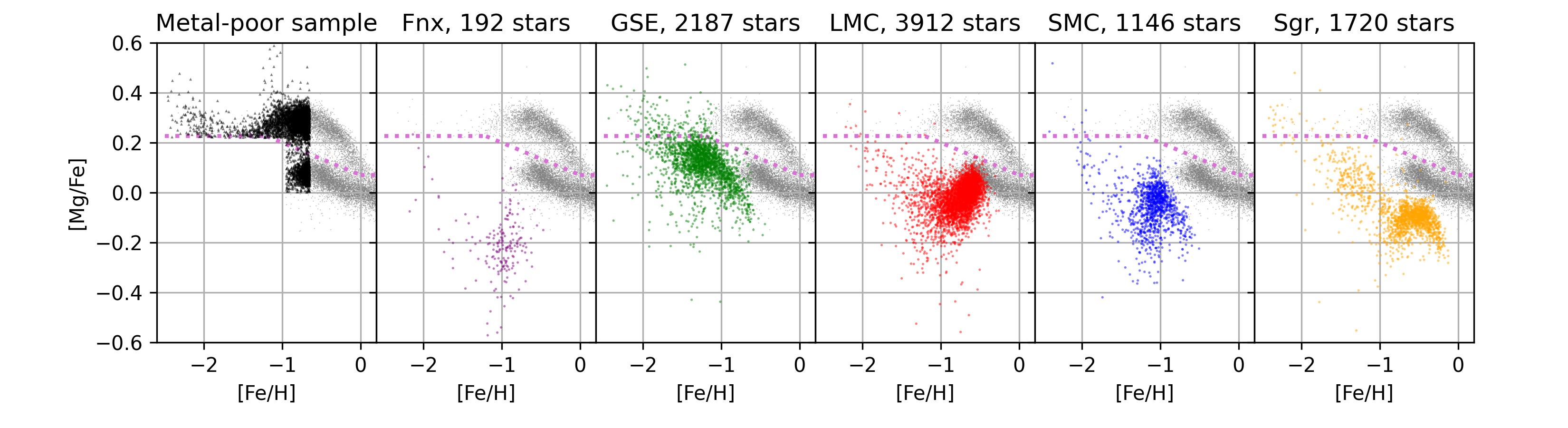}
    \caption{[Mg/Fe] vs. [Fe/H] for the metal-poor MW sample (described in the text) and the sample for each of the satellite galaxies \citep[the latter follows][]{Hasselquist2021}. The MW T$_{\rm eff}$-matched sample is reproduced in each panel as grey points. The number of stars in each sample is indicated in the title of each plot.\label{fig:mg_fe_all}. A global [Mg/Fe] offset of -0.05 has been applied such that MW stars at Solar [Fe/H] have [Mg/Fe] = 0.0.}
\end{figure*}

The stellar parameter distributions of these stars are shown in the red hatched histograms of Figure \ref{fig:param_select}. To demonstrate exactly which stars we are including at these low metallicities, we show in Figure \ref{fig:mg_fe_all} the [Mg/Fe]-[Fe/H] abundance plane for this MW metal-poor sample alongside each satellite galaxy sample. A global [Mg/Fe] offset of -0.05 is applied such that stars with solar [Fe/H] have [Mg/Fe] = 0.0. The left panel of Figure \ref{fig:mg_fe_all} shows that the metal-poor MW sample at [Mg/Fe] $>$ 0.27 likely contains some amount of GSE contamination, particularly in the metallicity range -2.0 $<$ [Fe/H] $<$ -1.5. However, as long as the ``pure CCSNe'' abundances of a GSE star do not differ significantly from those of a ``pure CCSNe'' MW star, this should have a negligible effect on our results. It is more important that the metal poor end of the MW low-$\alpha$ sequence does not contain GSE contamination, as the most metal-rich GSE stars have a much lower [Mg/Fe] abundance than the most metal-poor MW stars (as shown in the third from left panel of Figure \ref{fig:mg_fe_all}), which would cause us to fit 2-process model parameters to a median that is perhaps in between these sequences (described in more detail in Figure \ref{sec:results_vectors}). Moreover, as described in Section \ref{sec:resid_abund_results}, our results that rely on the MW 2-process model are limited to the metallicity range -1.0 $<$ [Mg/H] $<$ 0.0, where the high-Ia MW sample overlaps with the satellite galaxies, and are thus not affected by potential GSE contamination. A more in-depth understanding of how the 2-process model parameters and subsequent predictions are affected by the GSE/MW ambiguity at [Fe/H] $<$ -1.5 can be found in Griffith et al. in prep.

\section{The 2-process Model} \label{sec:two_proc}

The 2-process model (described in detail in \citealt{Weinberg2019} and \citealt{Weinberg2022}) is a tool to predict the chemical abundances of a star given its [Mg/Fe] and [Mg/H] abundances. The model assumes that the chemical abundance pattern of a star is a result of chemical enrichment from a prompt enrichment process and a delayed enrichment process, which can be physically linked to SNe II and SNe Ia, respectively. We note that the model is agnostic to what or how many physical mechanisms actually comprise the prompt and delayed enrichment processes. To ``train'' the model, one uses median abundance trends of low-Ia (a.k.a. high-$\alpha$) and high-Ia (a.k.a. low-$\alpha$) stars to derive two ``process vectors'' ($q_{\rm cc}$ and $q_{\rm Ia}$), which describe the IMF-averaged yields of each element within each process at a given metallicity. We derive these vectors in bins of size 0.1 dex in [Mg/H]. The predicted abundances of a given star depend on the amplitudes (A$_{\rm cc}$ and A$_{\rm Ia}$) of these two processes, which can be inferred from Mg and Fe alone or by fitting multiple elements. For detailed formulations of the model and its assumptions we refer the reader to \citet{Weinberg2022} and \citet{Griffith2024}.

The goals of this work are (1) to understand the extent to which a 2-process model with parameters fit to the MW abundance patterns is able to predict the abundances of other galaxies and (2) to understand the extent to which a 2-process model with parameters fit to a single dwarf galaxy (we choose Sgr) is able to predict the abundances of the other satellite galaxies. We therefore derive two sets of abundance vectors for each scenario. Inferring the model parameters by solving the abundance vector equations requires us to define two separate groups with distinct levels of Type Ia SNe enrichment at a given level of CCSNe enrichment. The division we use to define the ``low-Ia'' group is:

\begin{itemize}
    \item $[$Fe/H$]$  $>$ -1.2 \& [Mg/Fe] $>$ 0.12 - 0.13*[Fe/H] (same as \citealt{Weinberg2019}),
    \item $[$Fe/H$]$  $>$ 0.0 \& [Mg/Fe] $>$ 0.12 (same as \citealt{Weinberg2019}),
    \item $[$Fe/H$]$  $<$ -1.2 \& [Mg/Fe] $>$ 0.27 (introduced in this work).
\end{itemize}

We chose these divisions to be consistent with \citet{Weinberg2019} and \citet{Weinberg2022} where the metallicites of their sample overlaps with ours, but introduce a new [Mg/Fe] division at [Fe/H] $<$ -1.2 to select metal-poor stars that have [Mg/Fe] abundances close to the assumed ``pure'' prompt ratio of [Mg/Fe] = +0.3, described in detail below. Because we are ultimately fitting model parameters to the median trends of these samples, the exact placement of these divisions as well as whether we divide in the [Mg/Fe]-[Fe/H] abundance plane or the [Fe/Mg]-[Mg/H] abundance plane has little to no impact on the model parameters that we derive.

\begin{figure*}
    \plotone{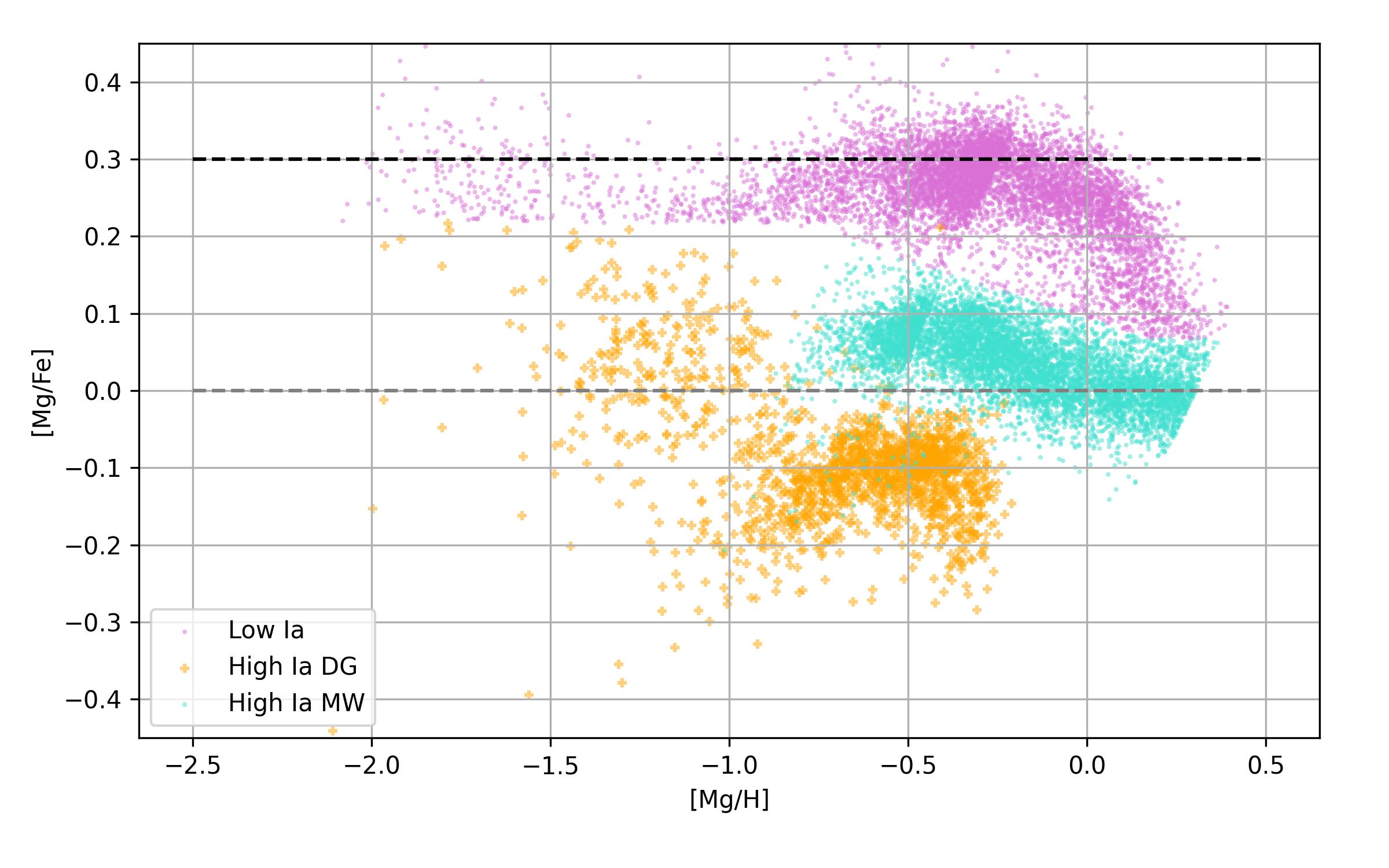}
    \caption{ [Mg/Fe]-[Mg/H] abundance patterns for the MW low-Ia sample (orchid), MW high-Ia sample (turquoise), and DG high-Ia sample (orange). \label{fig:2proc_init_init}. The black dashed horizontal line marks the assumed ``pure'' core-collapse Mg/Fe ratio of [Mg/Fe] = 0.3. The grey dashed line marks the solar Mg/Fe ratio of [Mg/Fe] = 0.0.}
\end{figure*}

This low-Ia sample, which is assumed to contain chemical enrichment primarily or entirely from the prompt process, is plotted as orchid-colored points in Figure \ref{fig:2proc_init_init}. For goal (1) listed above, the high-Ia sample used to infer the parameters is comprised of all MW stars selected in \autoref{sec:data} that have [Mg/Fe] abundances below the above division, plotted as turquoise-colored points in Figure \ref{fig:2proc_init_init} and henceforth referred to as the ``high-Ia MW'' sample. For goal (2) listed above, the high-Ia sample is instead comprised of all Sgr stars that have [Mg/Fe] below the above division. We refer to this sample as the ``high-Ia DG'' sample to indicate that it is used to fit the model parameters with the goal of comparing to other dwarf galaxies. We could have chosen any galaxy for goal (2), but decided to select Sgr because it spans the entire metallicity range of all satellite galaxies studied here, and, unlike the LMC, has a slightly more coherent [Mg/Fe]-[Fe/H] abundance pattern than the LMC at -2.0 $<$ [Fe/H] $<$ -1.0 (see e.g., Figure \ref{fig:mg_fe_all}). We infer separate sets of abundance vectors for using each high-Ia sample combined with the same low-Ia sample.

Because of small systematic differences in chemical abundances with stellar parameters, the APOGEE abundances have had zero point shifts applied to them to force stars at solar metallicity near the Sun to have solar abundances (see e.g., \citealt{Jonsson2020}). However, these offsets were applied to stars with generally hotter effective temperature than our sample here. Therefore, for each chemical element, we apply a new zero-point offset to force stars at solar [Mg/H] to have solar [X/Mg] abundances. These offsets are generally small ($<$ 0.05 dex), and are listed in Table \ref{tab:off}. Applying these offsets allows for a more direct comparison with \citet{Weinberg2022}. We then solve the 2-process equations to return the $q_{\rm cc}$ and $q_{\rm Ia}$ abundance vectors. As discussed in \citet{Weinberg2022}, these offsets have a small effect on the inferred abundance vectors, but have no effect on the residuals.

\begin{deluxetable}{l c}
\tablewidth{0pt}
\tablecolumns{12}
\tablecaption{Zero-point Offsets \label{tab:off}}
\tablehead{\colhead{\textbf{[X/Mg]}} & \colhead{\textbf{Offset (dex)}}}
\startdata
   O    &  0.018 \\
   Si   & -0.020 \\
   S    &  0.002 \\
   Ca   &  0.006 \\
   C+N  & -0.013 \\
   Na   & -0.003 \\
   Al   & -0.024 \\
   K    &  0.001 \\
   Cr   & -0.015 \\
   Ni   & -0.021 \\
   V    & -0.017 \\
   Mn   & -0.008 \\
   Co   &  0.003 \\
   Ce   & -0.004 \\
\enddata
\end{deluxetable}

\begin{figure*}
    \includegraphics[width=\textwidth]{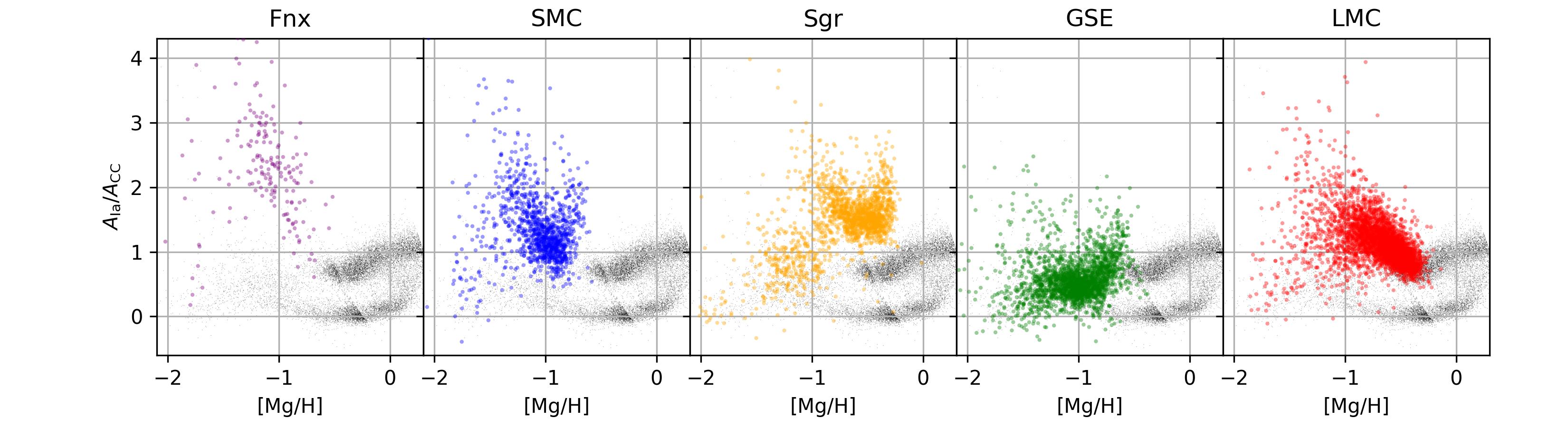}
    \caption{Ratios of the 2-process amplitudes, $A_{\mathrm{Ia}}/A_{\mathrm{CC}}$, for each galaxy plotted as function of metallicity ([Mg/H]). The MW sample is plotted in black and reproduced in each panel for comparison. Solar abundances correspond to $A_{\mathrm{Ia}} = A_{\mathrm{cc}} = 1$, so stars with a $A_{\mathrm{Ia}}/A_{\mathrm{cc}}$ ratio $>$ 1 have enhanced relative contribution from the ``delayed'' process, which we attribute to Type Ia SNe. \label{fig:2proc_ratio}}
\end{figure*}

Before showing the abundance vectors, we first show in Figure \ref{fig:2proc_ratio} the ratio of the delayed to prompt process for each star in each galaxy, as given by: $$A_{\mathrm{Ia}}/A_{\mathrm{cc}} = \frac{10^{\rm [Fe/Mg]} - 10^{\rm [Fe/Mg]_{pl}}}{1-10^{\rm [Fe/Mg]_{pl}}}$$ This is inferred directly from the [Fe/Mg] of a star assuming that the pure CCSNe-only  [Fe/Mg] abundance along the low-Ia/high-$\alpha$ ``plateau'' (${\rm [Fe/Mg]_{pl}}$, shown as the black dashed horizontal line in Figure \ref{fig:2proc_init_init}) is -0.3, as done in \citealt{Weinberg2019}. This $A_{\mathrm{Ia}}/A_{\mathrm{cc}}$ ratio plotted as a function of metallicity is useful to understand which process is contributing most significantly to the abundance of an element at a given point of a galaxy's evolution, providing insight into the general star formation history of a galaxy. Unlike \citet{Weinberg2022} and \citet{Sit2024}, but like \citet{Weinberg2019}, we infer this ratio directly from the [Fe/Mg] values alone rather than the iterative multi-element fit. We similarly infer the amplitude $A_{\mathrm{cc}}$ directly from [Mg/H] using $A_{\mathrm{cc}} = 10^{\mathrm{[Mg/H]}}$, which assumes Mg to arise purely from the prompt (CCSNe) process and normalizes to $A_{\mathrm{cc}} = 1$ at solar [Mg/H].   

As shown in the left panel of Figure \ref{fig:2proc_ratio}, Fornax (Fnx) stars generally have higher $A_{\mathrm{Ia}}/A_{\mathrm{cc}}$ ratios than the stars in the other 4 satellite galaxies, indicating that this galaxy has had the largest relative contribution of the delayed process to its evolution. GSE exhibits a relatively simple $A_{\mathrm{Ia}}/A_{\mathrm{cc}}$-[Mg/H] pattern, with a low $A_{\mathrm{Ia}}/A_{\mathrm{cc}}$ ratio that stays relatively flat at low metallicity during which GSE formed stars from primarily gas polluted by the prompt core-collapse process alone. At [Mg/H] $>$ -0.8, the $A_{\mathrm{Ia}}/A_{\mathrm{cc}}$ increases as Type Ia SNe began to contribute to the evolution of GSE. Sgr, SMC, and LMC all show a similar initial increase of the $A_{\mathrm{Ia}}/A_{\mathrm{cc}}$ ratio with increasing metallicity at low metallicity followed by a rapid decrease at higher metallicity, which can be interpreted as a later burst of star formation (see also \citealt{Hendricks2014,Nidever2020,Hasselquist2021}. The SMC and Sgr then show a final increase in $A_{\mathrm{Ia}}/A_{\mathrm{cc}}$ ratio at the highest metallicities as the starbursts decline and Type Ia SNe begin to contribute anew to the chemical evolution of these galaxies. Only the LMC and Sgr overlap significantly with the high-Ia MW sample. At these metallicities, the LMC stars have about the same $A_{\mathrm{Ia}}/A_{\mathrm{cc}}$ ratio as the MW stars whereas the Sgr stars are about a factor of 2 higher. 

\section{Abundance Vectors of the 2-process Model} \label{sec:results_vectors}

\begin{figure*}
\plotone{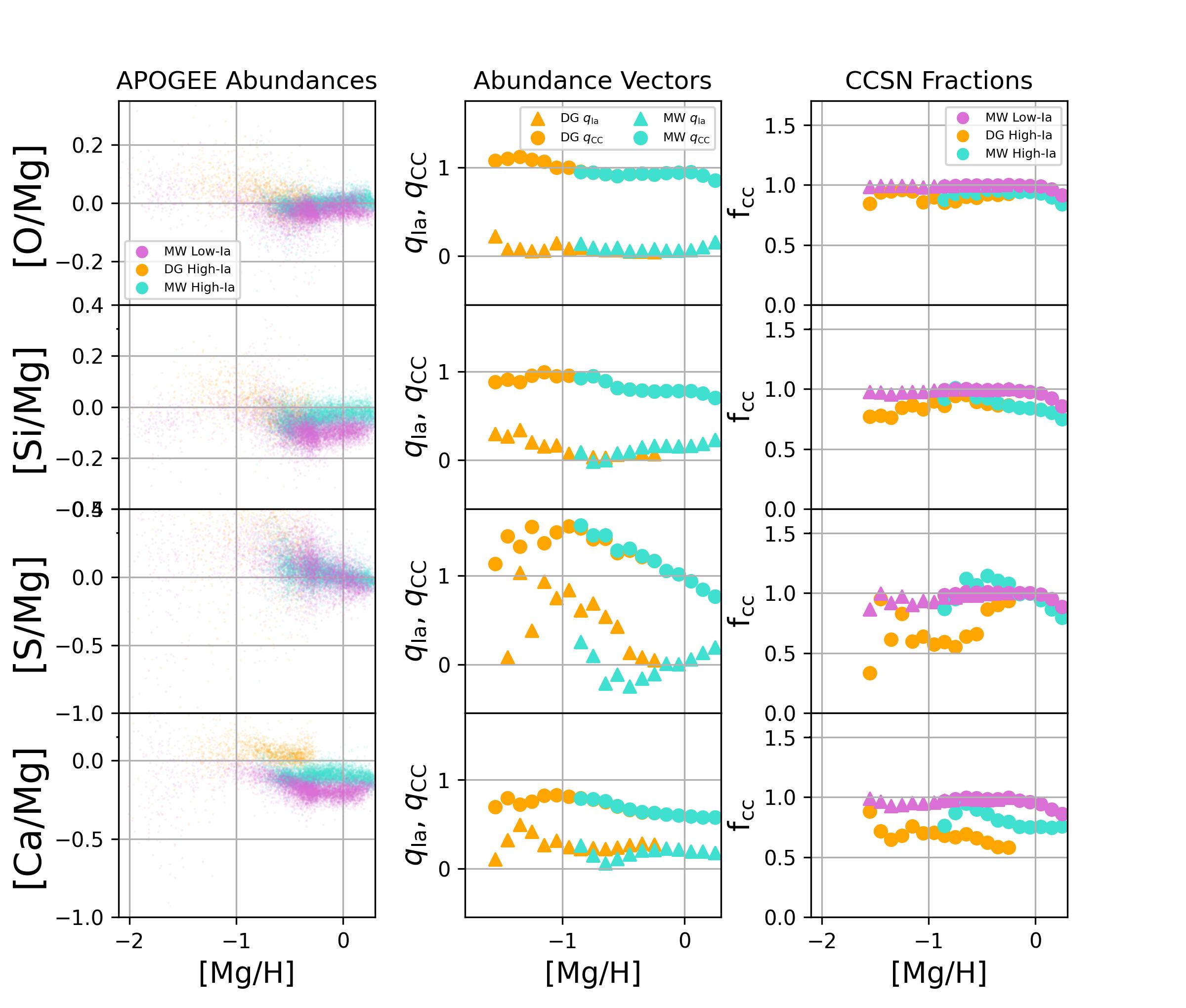}
\caption{Left column: APOGEE $\alpha$-element abundances for the high-Ia DG (orange), high-Ia MW (turquoise), and low-Ia (orchid) samples used to derive the parameters of the 2-process Model. The middle column shows the abundance vectors for each element in orange (fit to high-Ia DG sample) and turqoise (fit to high-Ia MW sample). The right column shows f$_{cc}$ value for each sample, which indicates the fraction of the element that comes from the prompt core-collapse SNe component at that metallicity for stars along the median low-Ia or high-Ia track of the MW or the median high-Ia track of the DG sample. \label{fig:2proc_init_1}}
\end{figure*}

\begin{figure*}
\plotone{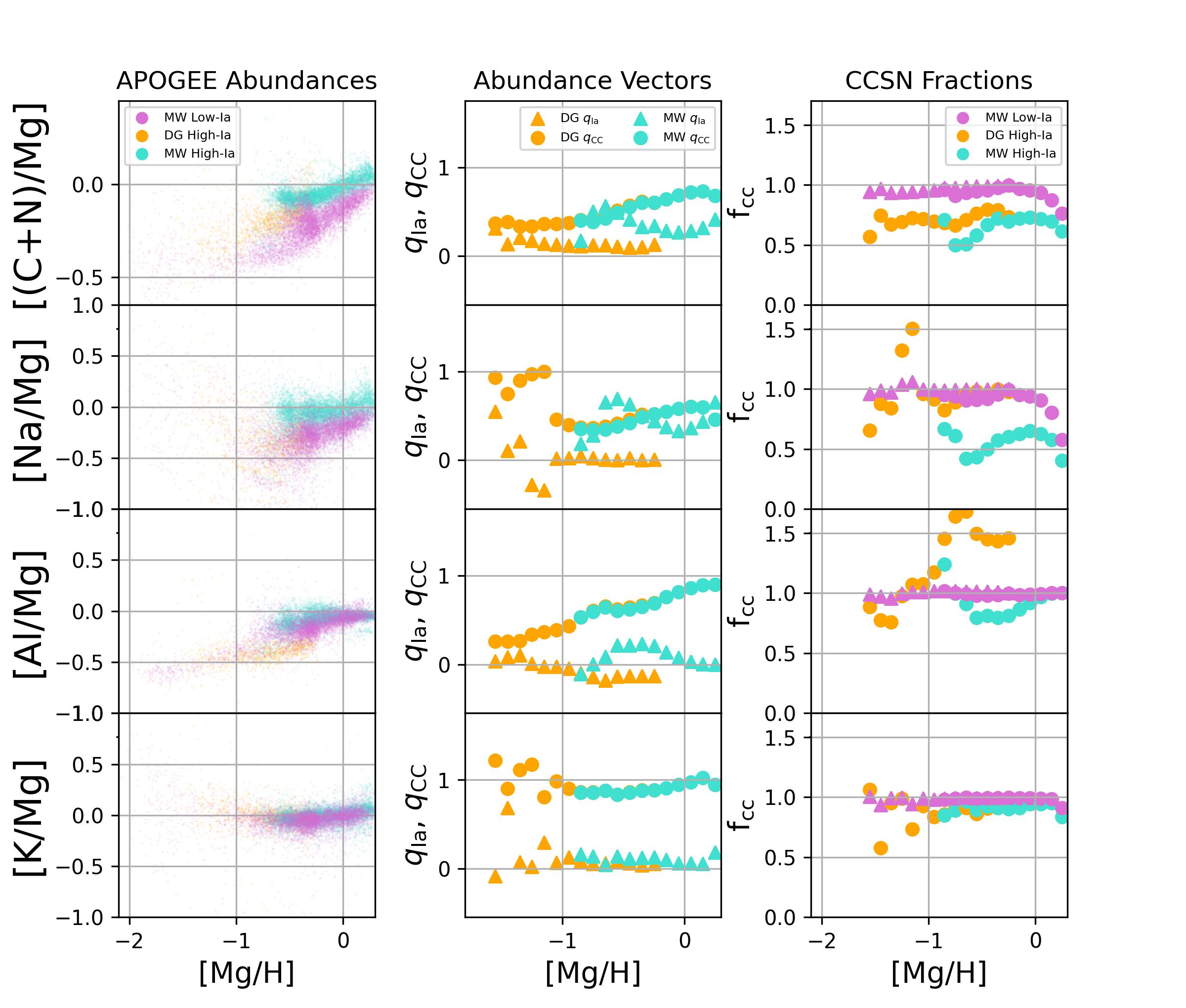}
\caption{Same as Figure \ref{fig:2proc_init_1}, but for the Odd-Z elements. \label{fig:2proc_init_2}}
\end{figure*}

\begin{figure*}
\plotone{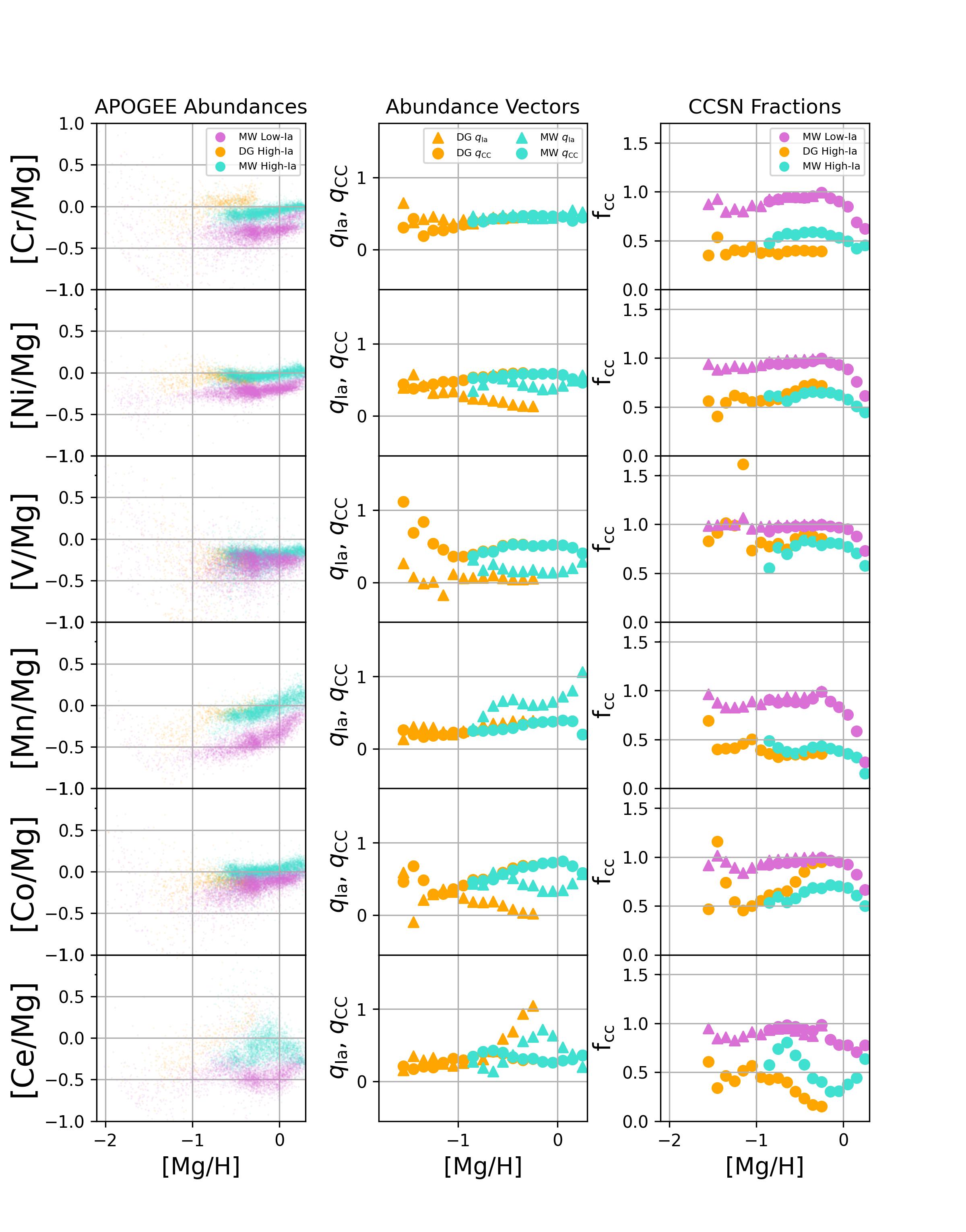}
\caption{Same as Figure \ref{fig:2proc_init_1}, but for the Fe-Peak$^{+}$ elements. \label{fig:2proc_init_3}}
\end{figure*}

Using the samples described in Section \ref{sec:data} and shown in Figure \ref{fig:2proc_init_init}, we derive two sets of abundance vectors. For display purposes and to assist in the discussion, we group the elements into 3 groups:

\begin{enumerate}
    \item {\textbf{$\alpha$ elements}:} O, Si, S, and Ca (\autoref{fig:2proc_init_1})
    \item {\textbf{Odd-Z elements\footnote{C is not an Odd-Z element, but due to stellar evolution effects, we cannot consider C and N separately as their surface abundances are affected by mixing mechanisms in red giant branch stars \citep{Salaris2015}. We instead consider the number-weighted sum of C and N abundances, as the sum of these elements after mixing events should be similar to the birth C+N abundance }}:} C+N, Na, Al, and K (\autoref{fig:2proc_init_2})
    \item {\textbf{Fe-Peak$^{+}$}:} Cr, Ni, V, Mn, Co, and Ce\footnote{ Ce is commonly considered an $s$-process element and is synthesized in large quantities by AGB stars, but we include with the heaviest elements here for presentation purposes.} (\autoref{fig:2proc_init_3}).
\end{enumerate}

The elemental abundances of each sample are shown in the left panels of Figures \ref{fig:2proc_init_1} - \ref{fig:2proc_init_3} and the inferred 2-process abundance vectors are shown in the middle panels. We also show in the right panels the f$_{\rm cc}$ values for each element along each sample, defined as:

$$\rm{f}_{\rm cc} = \frac{A_{\mathrm{cc}}q_{\rm cc}}{A_{\mathrm{cc}}q_{\rm cc} + A_{\mathrm{Ia}}q_{\rm Ia}}$$

This f$_{\rm cc}$ value indicates how much of some element comes from the ``prompt'' (assumed to be core-collapse) enrichment mechanism, and can therefore be a useful tool in understanding the dominant enrichment mechanism at a given metallicity for each stellar sample. f$_{\rm cc}$ $\simeq$ 1 indicates that element is being produced entirely in core-collapse SNe and f$_{\rm cc} < 0.5$ indicates the delayed mechanism (e.g., Type Ia SNe) is the dominant enrichment mechanism. Note that even if an element has a large IMF-averaged Type Ia SNe yield as indicated by the $q_{\rm Ia}$ vector  (e.g., Ni, with $q_{\rm Ia} \approx 0.5$), f$_{\rm cc} \approx 1$ for low-Ia stars (orchid points in Figures \ref{fig:2proc_init_1} - \ref{fig:2proc_init_3}) because these stars are inferred to have nearly pure CC SNe enrichment based on their [Fe/Mg] abundances. Additionally, because the f$_{\rm cc}$ value depends on the abundance vectors, they can be different for the high-Ia DG sample and the high-Ia MW sample, as we infer different abundance vectors for these samples for some elements (e.g., Ca, C+N, Al, Ni, and Ce, described in more detail below). We find that the abundance vectors we infer from the high-Ia MW sample are nearly the same as those inferred by \citet{Weinberg2022}, despite the slightly lower T$_{\rm eff}$ and $\log{g}$ distribution of our high-Ia MW sample.

In the following section, we use these two sets of abundance vectors to predict the elemental abundances of each galaxy, focusing on what the residuals might mean for galaxy evolution or breakdowns in the 2-process model assumptions. However, the fact that we even derive different abundance vectors for some elements between the high-Ia MW sample and the high-Ia DG sample (Sgr) suggests we should expect large residuals when using the high-Ia MW sample to predict the abundances of the various satellite galaxies. In the context of the 2-process model, different abundance vectors mean that the IMF-averaged yields of that process are different between the MW and Sgr. For elements like C+N (top row of Figure \ref{fig:2proc_init_2} or Ce (bottom row of Figure \ref{fig:2proc_init_3}, the differences in the inferred $q_{\rm Ia}$ vectors could be related to an increased contribution of AGB stars to the delayed process in Sgr. The causes for differences in the inferred $q_{\rm Ia}$ vectors for elements like Ni and Co (Figure \ref{fig:2proc_init_3}) are less obvious. The small difference in $q_{\rm Ia}$ vectors for Ca (Figure \ref{fig:2proc_init_1}) result in large changes in inferred f$_{\rm cc}$ values, suggesting the delayed process contributed at a much larger fraction to the Ca in Sgr than it did in the MW (if the model assumptions are correct). Finally, we actually infer negative $q_{\rm Ia}$ vectors for Al, leading to nonphysical  values for f$_{\rm cc}$ (f$_{\rm cc}$ $>$ 1), suggesting there is likely some issue with the 2-process model assumptions for this element. We explore potential causes for these differences of inferred $q_{\rm Ia}$ in Section \ref{sec:disc}, but first we analyze how well each model can predict the abundances of stars across all galaxies.

\section{Residual Abundance Analysis Results}
\label{sec:resid_abund_results}

\begin{figure*}
    \includegraphics[width=\textwidth]{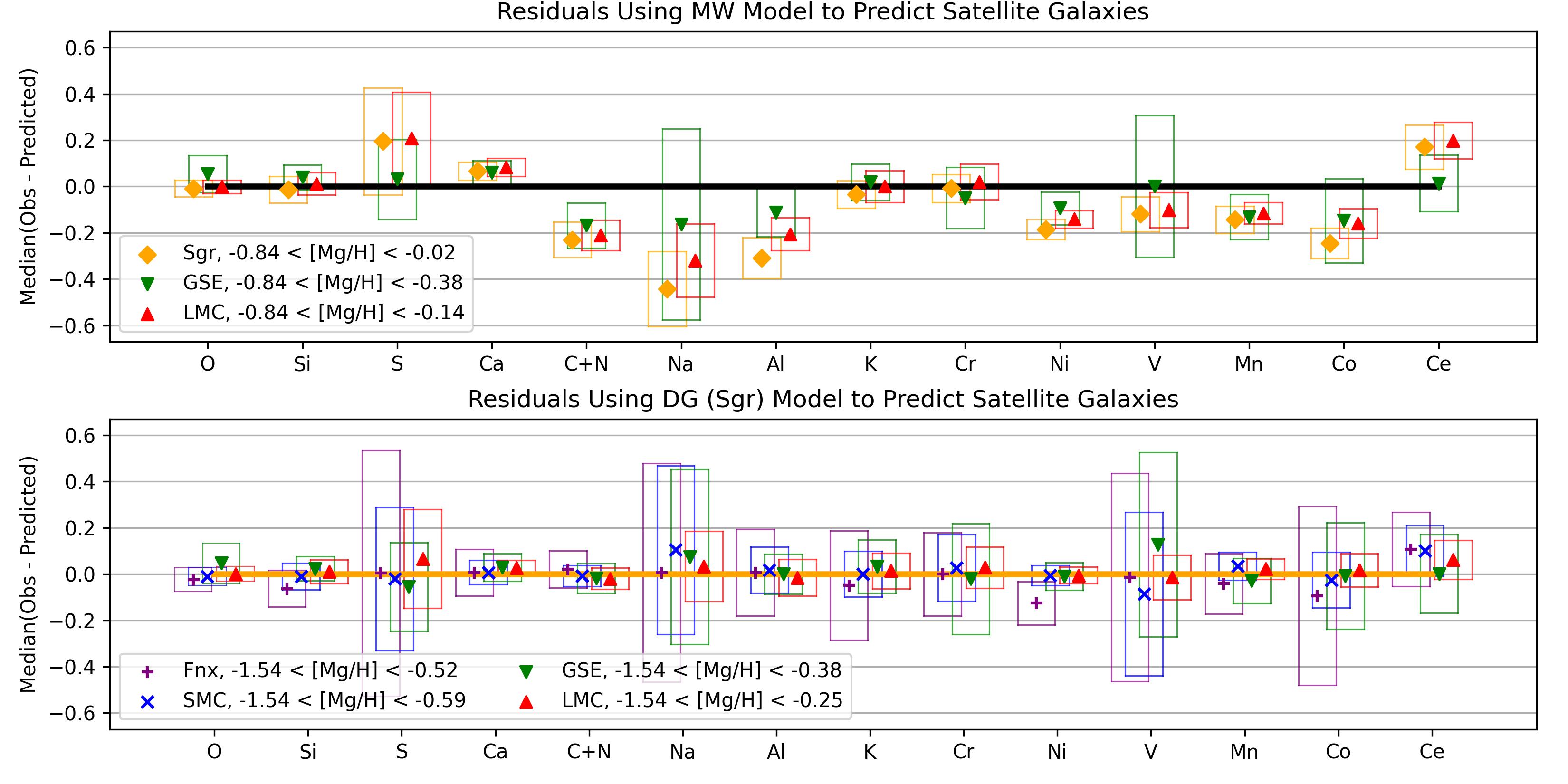}
    \caption{Median abundance residuals (observed - model prediction) for the satellite galaxy stars predicted using the 2-process model fit to MW stars (top panel) and for the satellite galaxy stars predicted using the 2-process model fit to the DG (Sgr) stars (bottom panel). The thick solid lines indicate where the residuals are 0, by construction, for each galaxy used to predict the model. The legends in the upper right indicate the metallicity range over which the residuals were calculated. The thinner solid lines around each point indicate the $\pm 1\sigma_{\rm MAD}$ (a robust measurement of the standard deviation using astropy.stats.mad\_std, $\sigma_{\rm MAD} \simeq$ 1.48*Median Absolute Deviation) of the distribution from which the median is taken, colored by galaxy, as indicated in the legend. \label{fig:resid_mw_sum}}
\end{figure*}

We use the derived abundance vectors shown in the middle panels of Figures \ref{fig:2proc_init_1} - \ref{fig:2proc_init_3} to predict the abundances of satellite galaxy stars. Because the high-Ia samples differ in metallicity coverage, the MW 2-process model is limited to predicting the abundances of the most metal-rich stars in the satellite galaxies, whereas the DG 2-process model is able to predict abundances for most of the metallicity range spanned by each satellite galaxy. In Figure \ref{fig:resid_mw_sum} we show the median abundance residuals (observed - predicted from the 2-process model) across all stars able to be predicted by each model for the vectors derived from the MW stars (top panel) and DG (Sgr) sample (bottom panel). The boxes indicate the $\pm 1\sigma_{\rm MAD}$ (a robust measurement of the standard deviation using astropy.stats.mad\_std, $\sigma_{\rm MAD} \simeq$ 1.48*Median Absolute Deviation) of the distribution from which the median is taken. The residuals of the satellite galaxies are much larger when the MW 2-process model is used to predict the abundances than when the DG 2-process model is used. Specifically, we find that, when using the MW model to predict satellite galaxy abundances:
\begin{itemize}
    \item The abundance of Ce is under-predicted by 0.2 dex for Sgr and the LMC, but not for GSE.
    \item The abundances of C+N, Na, Al, Ni, Mn, and Co are over-predicted by the model by 0.1 -- 0.3 dex, with the largest over-predictions found for Sgr and the LMC.
    \item The abundance of Ca is under-predicted by 0.1 dex for Sgr, LMC, and GSE, which is significant given the small scatter for each galaxy.
    \item The abundances of O, Si, K, and Cr are all close to the model predictions.
    \item S and V are the least well-measured abundances in our sample (as indicated by the large boxes that indicate scatter in the residuals in Figure \ref{fig:resid_mw_sum}), but the satellite galaxies show large residuals in both elements, with GSE showing the smallest residuals.
\end{itemize}

These residual abundance patterns are broadly consistent with what has been found in \citet{Weinberg2022} for a smaller sample of LMC stars and \citet{Sit2024} for the LMC and Sgr, the latter study carefully correcting for small systematic abundance differences as a function of $\log{g}$. In the context of the 2-process model, elements that are over-predicted by the model (negative values of observed - predicted abundances in the top panel Figure \ref{fig:resid_mw_sum}; e.g., C+N, Al, and Ni) mean that one or both of the processes are not producing enough of that element in a given galaxy given its stellar [Fe/Mg]-[Mg/H] abundances. We explore potential reasons for this throughout Section \ref{sec:disc}. Conversely, elements that are under-predicted by the model (positive values of observed - predicted abundances in the top panel of Figure \ref{fig:resid_mw_sum}; e.g., Ce) mean that that one or both of the processes are producing more of this element than is expected given the stellar [Fe/Mg]-[Mg/H] abundances, or that there is a separate process that is producing this element that is not active in the stellar sample used to fit the model parameters. In Section \ref{sec:agb}, we argue that the under-prediction of Ce in the satellite galaxies is due to an increased contribution from asymptotic giant branch (AGB) stars.

Despite the large residuals when using the MW 2-process model to predict the abundances of the satellite galaxies, the bottom panel of Figure \ref{fig:resid_mw_sum} shows that when using the DG 2-process model to predict the abundances of the satellite galaxies, the residuals are very close to zero for all galaxies, with differences up to 0.1 dex in a few cases. Therefore, the $q_{\rm cc}$ and $q_{\rm Ia}$ vectors that were fit to the DG sample changed in such a way relative to what they were when fit to the MW sample to accurately predict the abundances of the satellite galaxy stars. Specifically, the $q_{\rm Ia}$ vectors of C+N, Na, Al, Ni, V, and Co all decreased, whereas the $q_{\rm Ia}$ vectors of Ca and Ce increased, as seen by comparing the orange triangles to the turquoise triangles in the middle panels of Figures \ref{fig:2proc_init_1} - \ref{fig:2proc_init_3}. The changes in abundance vectors were able to accurately predict the satellite galaxy abundances despite the wide range in  $A_{\mathrm{Ia}}/A_{\mathrm{cc}}$ ratios (Figure \ref{fig:2proc_ratio}). 

The difference between the two panels of Figure \ref{fig:resid_mw_sum} is the principal result of this paper. In the range of overlapping metallicity, the APOGEE abundances of satellite galaxies stars are significantly different from those of the MW disk stars matched in [Mg/H] and [Mg/Fe], an effect seen for more than half of the elements examined here. However, once matched in [Mg/H] and [Mg/Fe], the abundances of these 5 satellite galaxies are quite similar to each other.

\section{Discussion}
\label{sec:disc}

The small residuals between observed and predicted abundances of the satellite galaxies when using the DG 2-process model suggest that the 2-process model is still a viable description of how chemical enrichment is occurring in the satellite galaxies, i.e., the amount of enrichment from the prompt source, [Mg/H], combined with the ratio of delayed-to-prompt enrichment ([Fe/Mg]), allows for accurate predictions of the abundances of all of the other elements. In the following discussion, we attempt to explain why there are large residuals for the satellite galaxies when using the MW 2-process model, linking the differences to potential differences in the enrichment processes or breakdowns in the model assumptions. We consider the following:

\begin{itemize}
    \item The delayed process, represented by the $q_{\rm Ia}$ abundance vector, is likely the sum of both Type Ia SNe ejecta and AGB ejecta, the latter of which likely contributes at a larger relative amount to the satellite galaxies than in the MW. This could explain the Ce under-predictions in the satellite galaxies using the MW 2-process model, and some of the C+N residual patterns (Section \ref{sec:agb}).
    \item The abundance vectors for the 2-process model are derived in bins of size 0.1 dex in [Mg/H], and thus any metallicity-dependent production of the elements is assumed to depend on Mg abundance. It is therefore possible that the model is not capturing the full intricacies of the metallicity-dependent production of some elements in one or both of the processes. For example, we find some evidence that the production of some of the metallicity-dependent elements, e.g., Al, is more dependent on C+N, suggesting that the Mg assumption might be responsible for some of the residuals in the satellite galaxies when using the MW 2-process model to predict their abundances (Section \ref{sec:metal_depend}).
    \item For an element with a metallicity-dependent yield, large inflows of pristine gas can drive departures from the 2-process model by reducing [Mg/H] without changing abundance ratios. A major dilution event could have affected the MW between the formation of the low-Ia and high-Ia sequences \citep[e.g.,][]{Chiappini1997,Spitoni2019}. This scenario could explain why most of the metallicity-dependent elements show large residuals in the satellite galaxies and the metallicity-independent elements do not (Section \ref{sec:inflows}).
    \item The initial mass function (IMF) and/or nature of the outflows may have been fundamentally different between the satellite galaxies and the MW, with the satellite galaxies preferentially losing some elements as compared to the MW. This plausibly explains elements with negative abundance vectors, e.g., Al and Na, as shown in the middle panel of Figure \ref{fig:2proc_init_2} (Section \ref{sec:imf}).
    \item The satellite galaxies may have had a larger fraction of atypical SNe, e.g.,  Sub-Chandrasekhar mass Type Ia SNe, to give rise to the larger residuals in Ca, Ni, Mn, and Co, all elements with increased contribution from the delayed process in the MW (Section \ref{sec:sne}).
\end{itemize}

The scenarios mentioned above and discussed in more detail below apply to the metal-rich (-0.9 $<$ [Mg/H] $<$ 0.00) stars of the LMC, Sgr, and GSE. However, any explanation or combination of explanations must account for the MW 2-process model residual patterns of these satellite galaxies as well as the fact that the DG 2-process model applied to the satellite galaxies results in little to no residuals for all elements across all satellite galaxies.

\subsection{The Effects of AGB Stars}
\label{sec:agb}

The 2-process model assumes that the elements are produced in some combination of a prompt and a delayed mechanism, which are typically attributed to CCSNe and Type Ia SNe, respectively. However, AGB stars also contribute to the chemical enrichment of a galaxy, producing a significant fraction of the C and N in MW disk stars (e.g., \citealt{Kobayashi2020}) and the majority of the Ce (e.g., \citealt{Prantzos2020}). In principle, the contribution of AGB stars would be included in the 2-process abundance vector of the delayed enrichment mechanism, $q_{\rm Ia}$. However, if the ratio of AGB enrichment to Type Ia SNe enrichment is lower for the MW than it is for the dwarf galaxies, then we would expect the satellite galaxies to have under-predicted abundances from the MW 2-process model for these elements. 

The top panel of Figure \ref{fig:resid_mw_sum} shows that the metal-rich Sgr and LMC stars have their Ce abundances under-predicted by 0.2 dex. The under-prediction is not seen in the GSE stars, plausibly because these stars have lower $A_{\mathrm{Ia}}/A_{\mathrm{cc}}$ ratios than Sgr and the LMC, over most of this metallicity range (see Figure \ref{fig:2proc_ratio}). Ideally, one could develop a 3-process model to separate the AGB enhancement, but calibrating such a model requires at least one additional element with a known ratio of AGB/SNIa/CCSN enrichment (see e.g., \citealt{Griffith2022}). Still, $s$-process enhancement has been measured in these galaxies before (e.g., \citealt{McWilliam2013}), and the under-prediction of Ce by the MW 2-process model is very likely a result of not properly taking into account a third AGB process.

\begin{figure*}
\plotone{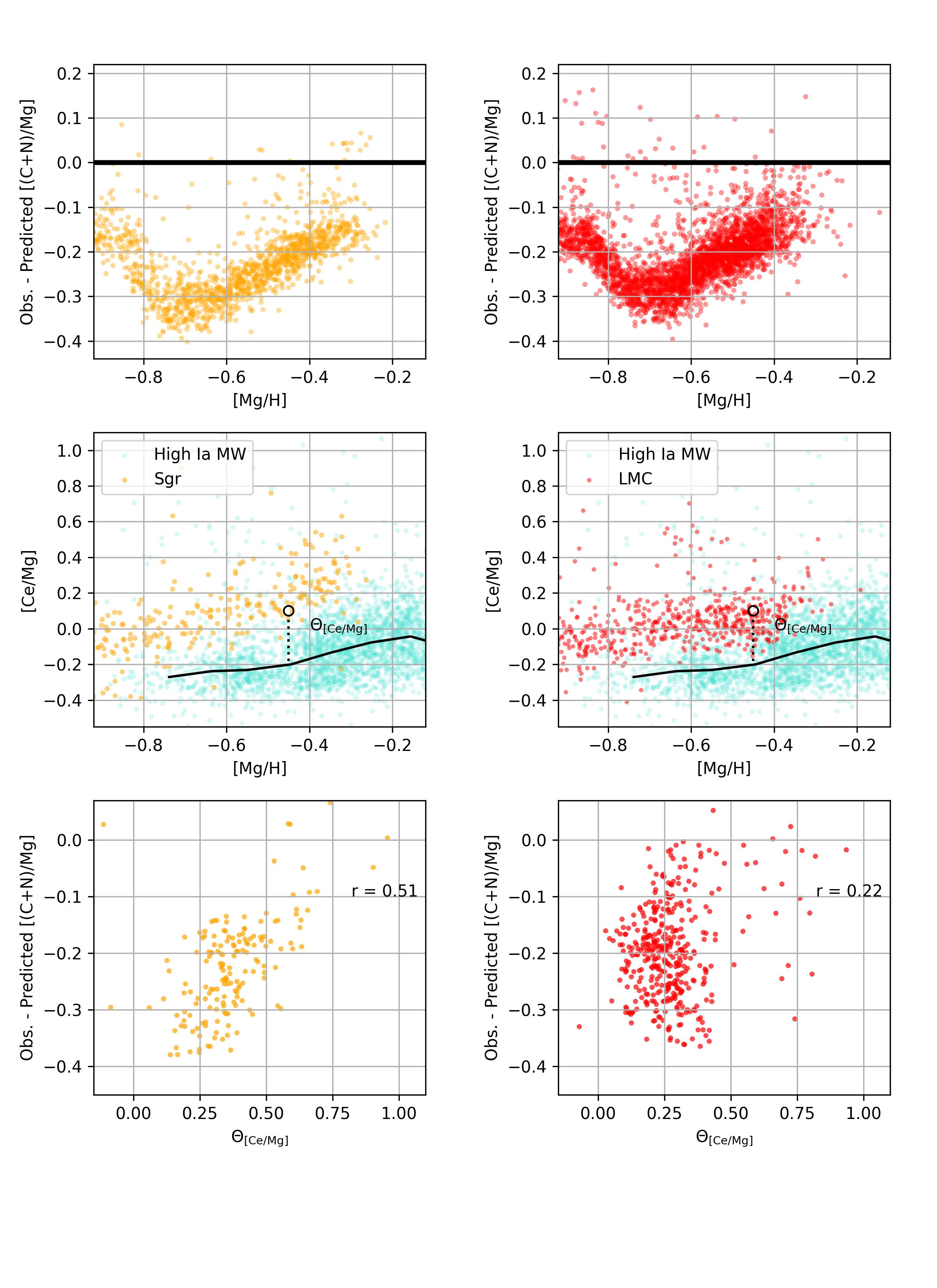}
\caption{Top: Residual [(C+N)/Mg] abundances as a function of [Mg/H] for Sgr (left) and the LMC (right). The thick black line marks the residual = 0.0 line. Middle: [Ce/Mg] abundance plotted as a function of [Mg/H] to demonstrate that the parameter $\Theta_{\rm [Ce/Mg]}$ is the excess [Ce/Mg] abundance of a satellite galaxy star above that of the MW median trend (black line). The high-Ia MW sample is reproduced in each panel as turquoise points. Bottom: the residual [(C+N)/Mg] abundance plotted as a function of excess [Ce/Mg], $\Theta_{\rm [Ce/Mg].}$ . Pearson correlation coefficients are indicated in the bottom panels.\label{fig:ce_check_resid}}
\end{figure*}

If the model under-predictions of [Ce/Mg] in the satellite galaxies at [Mg/H] $>$ -0.8 are due to a delayed process that has more AGB contribution than the delayed process in the MW, then we might also expect residual trends in [C/Mg] and [N/Mg] (which we show together as [(C+N)/Mg] to negate the effects of internal stellar mixing) to be correlated with [Ce/Mg], as these elements are also produced in non-negligible quantities in AGB stars \citep{Kobayashi2020}. In the top row of Figure \ref{fig:ce_check_resid}, we first show the residual [(C+N)/Mg] abundances as a function of [Mg/H] for both Sgr (left) and the LMC (right). We see that both galaxies, aside from the most metal-poor stars, exhibit a residual [(C+N)/Mg] abundance that gets closer to 0 with increasing [Mg/H]. The abundances of the most metal-rich stars are still over-predicted by $\sim$0.15 dex, suggesting that there are likely additional reasons beyond AGB enrichment to explain the [(C+N)/Mg] residuals.

Still, we can assess the extent to which AGB production might be influencing the C+N abundances of these galaxies by analyzing how the residuals change as a function of ``excess'' [Ce/Mg] abundance, which we notate as $\Theta_{\rm [Ce/Mg]}$. This quantity is defined as the difference between a star's [Ce/Mg] abundance and the abundance of the median high-Ia MW sample at that same metallicity, demonstrated in the middle panels of Figure \ref{fig:ce_check_resid}. Assuming that Ce is produced entirely in AGB stars at these metallicities, we use this as a measurement of excess AGB contribution in a star as compared to a ``typical'' star in the MW high-Ia sequence.

The bottom panel of Figure \ref{fig:ce_check_resid} shows that $\Theta_{\rm [Ce/Mg]}$ is correlated with the residual [(C+N)/Mg] abundance for Sgr, but not so for the LMC. So, for Sgr, the smaller C+N residuals at higher [Mg/H] are plausibly due to the fact that AGB stars begin to create C and N in large quantities, ``making up'' for whatever process failed to produce the expected amount of C+N from the MW 2-process model. This is apparently not the case for the LMC, but this might be expected given that the LMC has a much lower $A_{\mathrm{Ia}}/A_{\mathrm{cc}}$ ratio at these metallicities than Sgr (see Figure \ref{fig:2proc_ratio}). If AGB stars play a larger role in enriching the satellite galaxies than in the MW, we would expect to see the largest differences at metallicities where the delayed process dominates. However, AGB enrichment cannot explain why the C+N residual decreases with increasing [Mg/H] for the LMC, nor why these galaxies have such large residuals to begin with. 

In general, the [Ce/Mg] residuals of the satellite galaxies for the DG 2-process model are much smaller, as shown in the lower panel of Figure \ref{fig:resid_mw_sum}. The SMC and Fnx have [Ce/Mg] residuals of $\sim$ 0.1 dex, suggesting that these two galaxies have had the largest amount of AGB contribution to their chemical evolution.

\subsection{Metallicity Dependence}
\label{sec:metal_depend}
In addition to over-predicting [(C+N)/Mg], the MW 2-process model also over-predicts the abundances of [Al/Mg] and [Na/Mg] (see top panel of Figure \ref{fig:resid_mw_sum}). Aside from the aforementioned contributions of AGB stars to C+N, these elements are largely produced by the ``prompt'' process, and in such a way that depends on metallicity, as demonstrated by their abundance vectors shown in the middle panel of Figure \ref{fig:2proc_init_2}. One potential reason for these over-predictions could be that, while traditionally [Fe/H] has been used as a tracer for overall metal abundance (metallicity), the 2-process model uses [Mg/H] because it serves as a tracer for overall enrichment from the prompt process only (with the assumption that Mg is produced in the prompt process only). However, the detailed nucleosynthetic processes that actually create a specific element might be more dependent on other elements, such as C, N, and/or O, in the case of producing Na and Al via hydrostatic carbon and neon burning (see e.g., \citealt{Truran&Arnett1971}). Therefore, we might expect smaller Al and Na residuals if we use C+N as a reference element rather than Mg. 

\begin{figure*}
\plotone{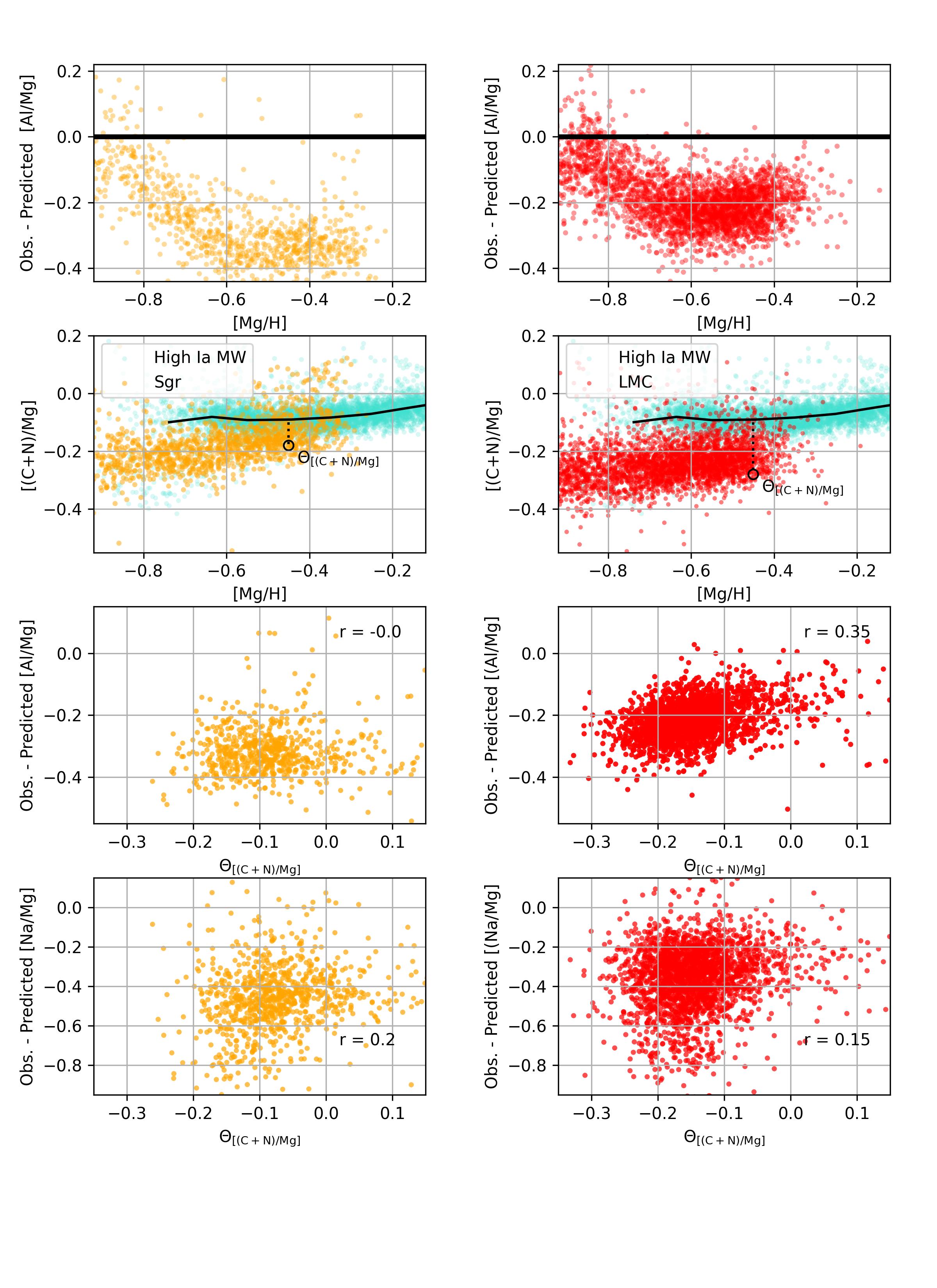}
\caption{Top Row: Residual [Al/Mg] abundances as a function of [Mg/H] for Sgr (left) and the LMC (right). The thick black line marks the residual = 0.0 line. Second Row: [(C+N)/Mg] abundance plotted as a function of [Mg/H] to demonstrate that the parameter $\Theta_{\rm [(C+N)/Mg]}$ is the excess [(C+N)/Mg] abundance of a satellite galaxy star above that of the MW median trend (black line). The high-Ia MW sample is reproduced in each panel as turquoise points. Third Row: the residual [Al/Mg] abundance plotted as a function of $\Theta_{\rm [(C+N)/Mg]}$. Fourth Row: the residual [Na/Mg] abundance plotted as a function of $\Theta_{\rm [(C+N)/Mg]}$. Pearson correlation coefficients are indicated in the bottom two rows.\label{fig:cn_check_resid}}
\end{figure*}

Unfortunately, it is not trivial to swap in C+N for Mg and re-derive the 2-process model parameters using C+N as the reference element, as C+N is created in both processes. Instead, similar to what was done in Figure \ref{fig:ce_check_resid}, we first look at whether or not the residual Al abundances are correlated with C+N abundance, which might indicate that the production of Al depends on the abundance of C+N. The top panel of Figure \ref{fig:cn_check_resid} shows [Al/Mg] residuals as a function of [Mg/H] again for Sgr (left) and the LMC (right). Sgr shows essentially flat residuals across all metallicity whereas the LMC shows residuals that get slightly closer to 0 with increasing [Mg/H], indicating that the more metal-rich stars in the LMC have [Al/Mg] abundances that are closer to the model predictions. The second row of Figure \ref{fig:cn_check_resid} introduces $\Theta_{\rm [(C+N)/Mg]}$, which is the difference between a star's [(C+N)/Mg] abundance and that of the median high-Ia MW trend. This is similar to $\Theta_{\rm [Ce/Mg]}$, described in Section \ref{sec:agb}, but for this element the satellite galaxies are deficient relative to the MW at fixed [Mg/H] instead of enhanced. The third row shows that this quantity is uncorrelated with [Al/Mg] abundance residuals for Sgr, but correlated with for the LMC. We do not see any correlations for Na, which might be because the Na abundances at these metallicities are too high.

\begin{figure*}
\plotone{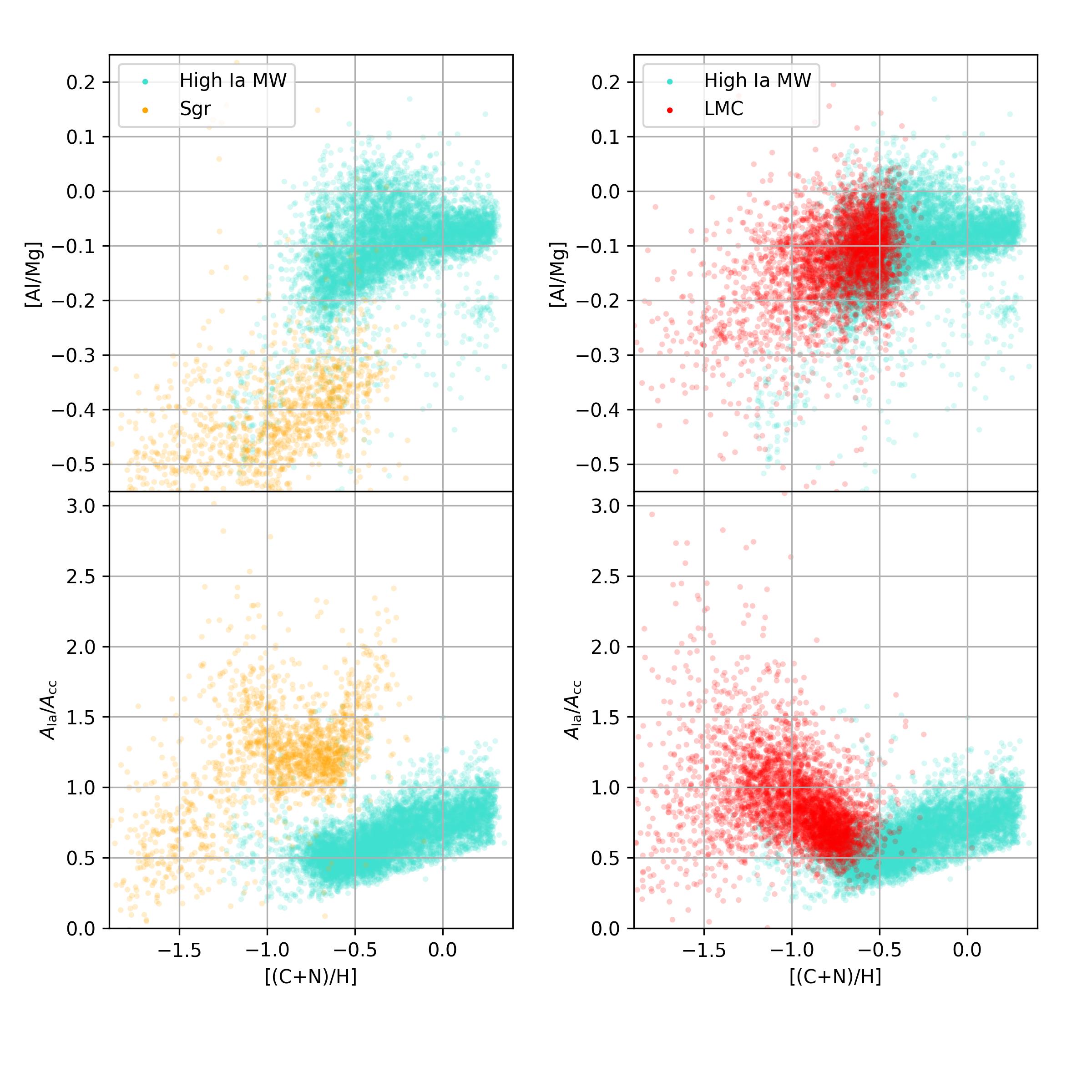}
\caption{Top: [Al/Mg] vs. [(C+N)/H] for Sgr (left, orange) and the LMC (right, red), with the MW high-Ia sample reproduced in each panel as the turquoise points. Bottom: $A_{\mathrm{Ia}}/A_{\mathrm{CC}}$ ratios as a function of [(C+N)/H]. \label{fig:cn_look}}
\end{figure*}

Thus, from the third row of Figure \ref{fig:cn_check_resid}, we again have a situation where one satellite galaxy exhibits a correlation and the other does not. However, we argued in Section \ref{sec:agb} that Sgr, with its higher $A_{\mathrm{Ia}}/A_{\mathrm{CC}}$ ratios and correlations between $\Theta_{\rm [Ce/Mg]}$ and residual [(C+N)/Mg], exhibits significant contribution from AGB stars to its C+N abundance at these metallicities. The LMC, on the other hand, seems to have C+N coming almost entirely from the prompt process at these metallicities. If the production of Al in the prompt process is more dependent on C+N than it is Mg, as theoretically expected, then we would expect the LMC to have more high-Ia MW-like [Al/Mg] abundances at fixed [(C+N)/H] abundance. This is exactly what we find in the upper-right panel of Figure \ref{fig:cn_look}. For the LMC stars, there is substantial overlap in [Al/Mg] with the MW high-Ia sample at fixed [(C+N)/H]. Sgr, shown in the upper-left panel of Figure \ref{fig:cn_look}, is still deficient, but that is plausibly because the C+N is coming from AGB for Sgr whereas it is almost entirely coming from the prompt process in the LMC, as indicated by the $A_{\mathrm{Ia}}/A_{\mathrm{CC}}$ ratios shown in the bottom panel of Figure \ref{fig:cn_look}.

This scenario doesn't quite explain why the [(C+N)/Mg] abundances of the satellite galaxies are low in the first place. However, it could explain why the DG 2-process model finds slightly negative $q_{\rm Ia}$ vectors for Al and Na, as shown in the middle panel of Figure \ref{fig:2proc_init_2}. Essentially the DG (Sgr) stars used to fit the model have a [Al/Mg] abundance that is lower than the MW low-Ia sample. At these metallicities, the [Al/Mg] abundance of the MW low-Ia sample represents the ``pure prompt'' abundance. To be below the pure prompt [Al/Mg] abundance means that the $q_{\rm Ia}$ vector must work to \emph{remove} Al from the galaxy, a description that is not necessarily physical. However, if a third process could be introduced to account for the increased [(C+N)/H] in Sgr due to AGB stars, one could, in principle, modify the 2-process model such that the metallicity dependence of elements like Al is on [(C+N)/H]$_{\rm cc}$ rather than [Mg/H]. Alternatively, we could fit the 2-process model to the LMC stars instead, and use [(C+N)/H] as the reference element as Section \ref{sec:agb} argues that the LMC has little AGB contribution to [(C+N)/H] at these metallicities. This is beyond the scope of this work, but we do emphasize that the upper-right panel of Figure \ref{fig:cn_look}, which shows the LMC stars overlapping with the high-Ia sample in [Al/Mg], is evidence for a stronger Al dependence on C+N than Mg in the prompt process.

\subsection{Inflows}
\label{sec:inflows}
While the [Al/Mg] overlap for the LMC and high-Ia MW samples at fixed [(C+N)/H] suggests that the assumed metallicity dependence of the model is a plausible explanation for the Al abundance residuals as well as the decreasing [(C+N)/Mg] residuals with increasing [Mg/H] for the LMC, this explanation still doesn't account for why the [(C+N)/Mg] abundances of the satellite galaxies are lower than the predictions to begin with. One potential explanation is that the inflow of pristine gas was a much larger factor in the chemical evolution of the MW than it was for the satellite galaxies. As described in detail below, the inflow of pristine gas would lower the overall metallicity of a galaxy while keeping the [X/Mg] abundance ratios at whatever level they were at right before the pristine inflow. For the metallicity-dependent elements, this would mean that stars formed immediately after the inflow would appear to have higher [X/Mg] ratios than their metallicity would suggest. This could lead to large residuals when using the MW 2-process model to predict the abundances in other galaxies that did not have such an inflow event.

A wide range of MW Galactic archaeological work over the past decades has put together a picture that the MW low-Ia (high-$\alpha$) sequence formed first during a period of vigorous star formation, with the gas forming stars enriching up to nearly solar metallicity before Type Ia SNe began to contribute to the evolution (e.g., \citealt{Tinsley1979}). The high-Ia sequence, with its more solar-like $\alpha$ element abundances, formed some time after. Moreover, this high-Ia sequence is only a sequence if stars from all Galactic disk positions are considered at the same time, with the outer disk containing a higher fraction of metal-poor stars and the inner disk containing a relatively higher fraction of metal-rich stars (e.g., \citealt{Hayden2015,Weinberg2019}). Age studies have revealed that, while each Galactic position contains a range of metallicities along the high-Ia sequence, the youngest stars become more metal-poor with increasing Galactic radius (e.g., \citealt{Mackereth2017,Hasselquist2019a}). Therefore, we are now putting together a picture of our Galaxy that shows that the low-Ia sequence really is an evolutionary sequence, and formed first. Then, after an infall of pristine gas (perhaps even from the GSE merger, e.g., \citealt{Ciuca2024}) that diluted the now super-solar gas of the MW to some metallicity, with this dilution apparently stronger with increasing Galactic radius (qualitatively consistent with \citealt{Buck2023} conclusions from simulations), stars began to form again at some point on the high-Ia sequence \citep{Chiappini1997,Spitoni2019}. Support for such a sudden infall was found by \citet{Spitoni2024}, who find that their two-infall model results in very few stars ($\sim$1\% of all stars formed) formed during this period of dilution. Radial migration then likely played a role in mixing high-Ia populations across the Galaxy (e.g., \citealt{Schonrich&Binney2009b,Hayden2015,Loebman2016,Hasselquist2019a,Lian2020d}), but what we care about in the 2-process model is that the MW high-Ia stars likely formed from a combination of gas that was enriched to super-solar metallicty and pristine gas.

\begin{figure*}
\plotone{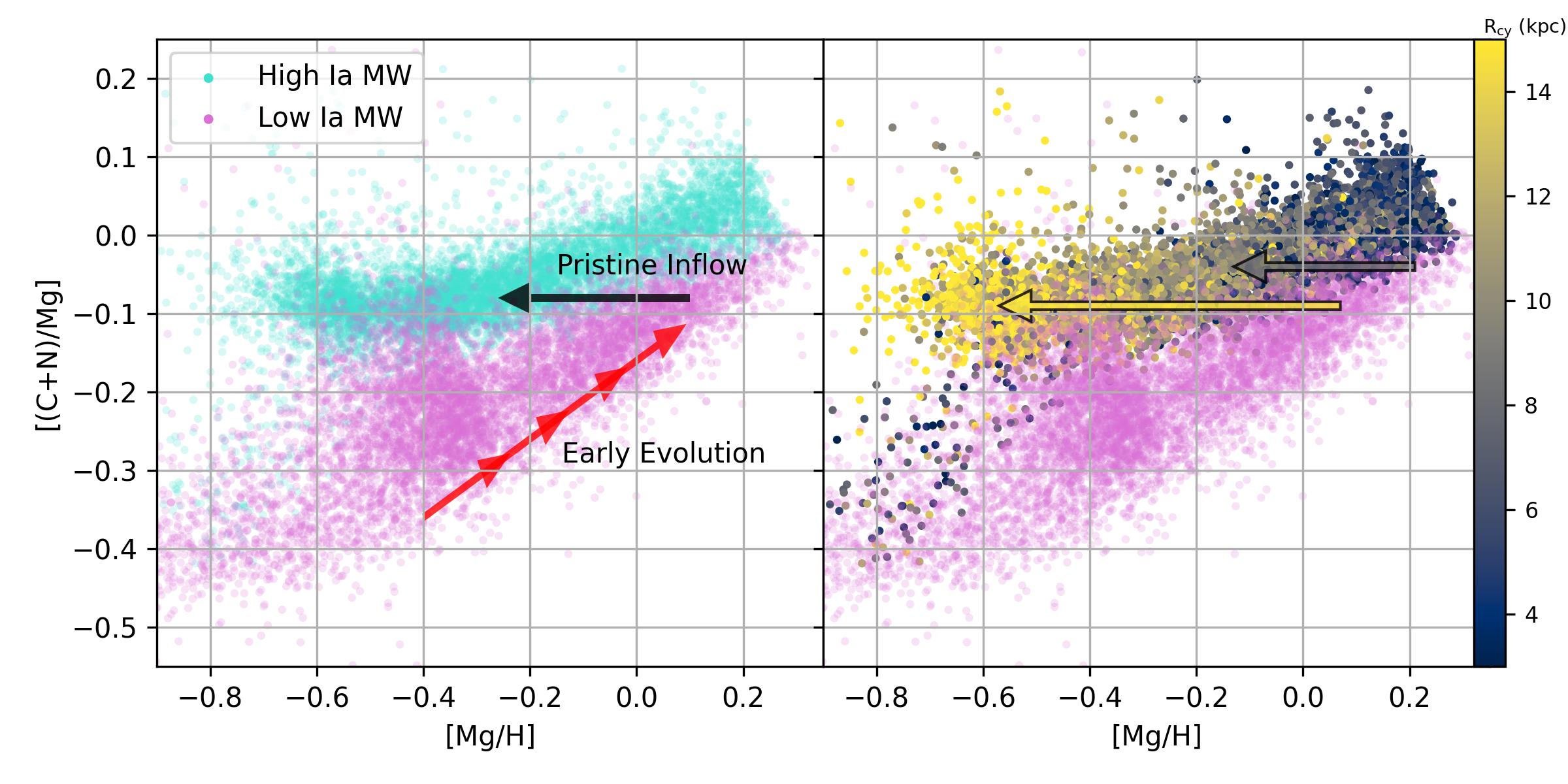}
\caption{Left: [(C+N)/Mg]-[Mg/H] abundance patterns for the high-Ia MW sample (turquoise) and the low-Ia MW sample (orchid). The series of red arrow indicates the time evolution of the low-Ia MW sample. The black arrow indicates how the abundance of gas changes when pristine gas is added to the MW (overall metallicity drops, but relative abundance stays the same). Right: same as left but now the high-Ia sample is colored by Galactic cylindrical radius (R$_{\rm cy}$). The arrows indicate that different regions of the MW potentially had more or less dilution before forming stars again.\label{fig:dilution}}
\end{figure*}

A schematic of this scenario is shown in Figure \ref{fig:dilution}, where we show the [(C+N)/Mg] - [Mg/H] abundance plane for just the two MW samples. C+N is chosen because it is an element that we have been discussing, although any metallicity-dependent element could be shown. In the left panel, the red arrow indicates the evolution along the low-Ia MW sequence, which formed in the early stages of the MW's history. The [(C+N)/Mg] abundance increases with increasing [Mg/H]. Then, at some point along this sequence, an infall of pristine gas would decrease the [Mg/H], but not change the [(C+N)/Mg] abundance, as indicated by the black arrow. The right panel shows the same idea, but now the high-Ia MW stars are colored by Galactic cylindrical radius ($R_{\rm cy}$), demonstrating that the high-Ia MW sequence is not necessarily a sequence in time, but can be a sequence that is populated by some combination of previously enriched gas mixed with some amount of pristine gas, the relative contribution of each apparently dependent on location in the Galaxy.

The point of this discussion is not necessarily to describe exactly what is happening in the MW, but to demonstrate that if there is a large amount of pristine inflow in the MW that occurs on relatively short timescales strong enough to significantly dilute ISM abundances, but this does not occur in the satellite galaxies, then we would expect the MW 2-process model to fail on predicting any metallicity-dependent element in the satellite galaxies. The elements that are predicted well for the satellite galaxies when using the MW 2-process model, as shown in the top panel of Figure  \ref{fig:resid_mw_sum}, tend to be elements that are not metallicity-dependent. A simple and fundamental difference such as pristine inflow would also explain why the DG 2-process model is able to accurately predict the abundances of the the satellite galaxies. However, given the different magnitude of the residuals from the MW 2-process model, it is not clear if pristine inflow can explain all of the residuals. Future work that adds a ``dilution'' vector to the 2-process model will help clarify the extent to which pristine inflow results in large residuals in the satellite galaxies when using the MW 2-process model.

\subsection{Outflows and IMF}
\label{sec:imf}

In addition to inflow of pristine gas, fundamental differences in outflows between the satellite galaxies and MW could explain some of the residuals, particularly the large under-predictions of C+N, Na, and Al. Being lower mass and, in some cases, in the process of disrupting, the satellite galaxies may not have been able to hold onto ejecta from SNe as effectively as the more massive MW. If such outflows were progenitor-dependent, where ejecta from more massive SNe were more likely to leave the system than less massive SNe, then we would expect deficiencies in elements that are produced in hydrostatic burning, the yields of which depend on the mass of the progenitor. Such differences in outflows would not be properly captured by the 2-process model, and would result in large residuals of elements like C+N, Na, and Al, which is what we observe.

\begin{figure*}
\includegraphics[width=\textwidth]{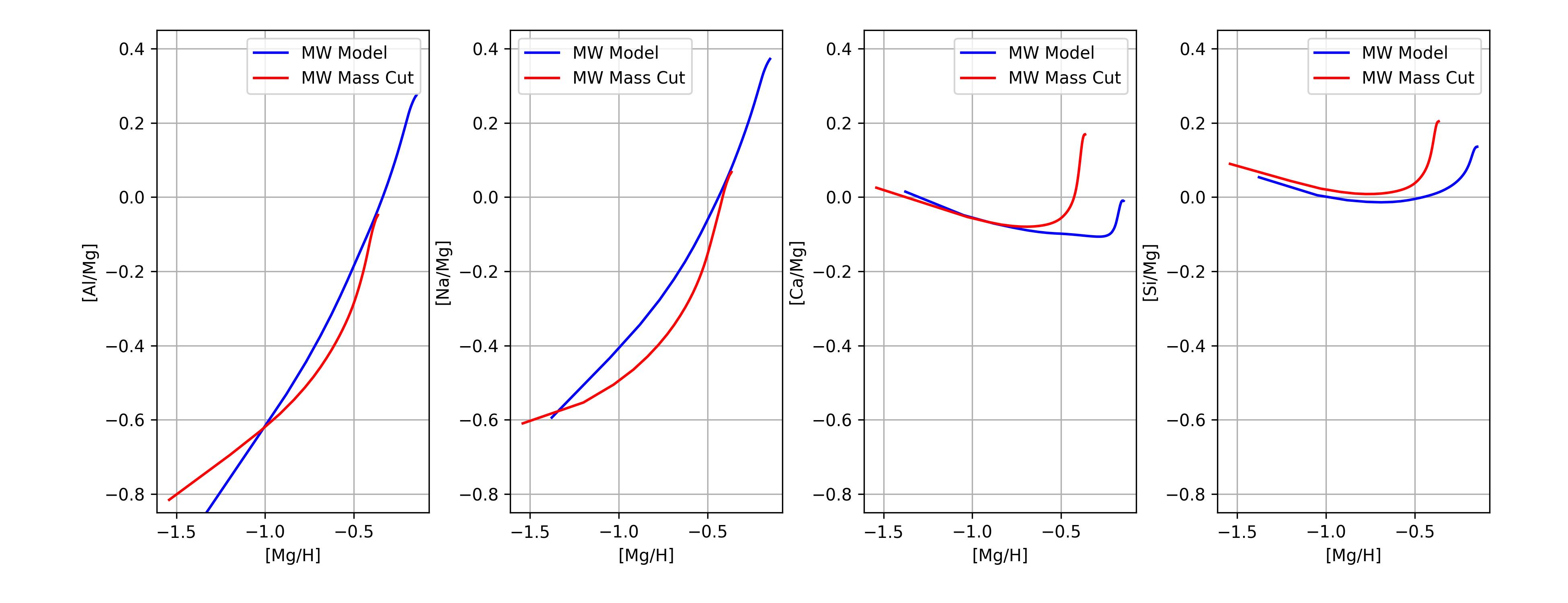}
\caption{Chemical evolution tracks in [X/Mg]-[Mg/H] abundance space for four elements. Chemical evolution tracks are generated using flexCE, with the blue lines showing the chemical tracks from a fiducial MW model and the red lines showing the chemical tracks from a model with the same parameters except for a mass cutoff of 30 solar masses such that stars beyond this mass do not form. \label{fig:almg}}
\end{figure*}

A different scenario, but with similar effects, is that star formation of the more recent populations of these stars occurred in a way that the most massive stars did not form, e.g., ``top-light'' IMF (e.g., \citealt{McWilliam2013,Hasselquist2017,Carlin2018}). If these massive stars never formed, then the mass-dependent hydrostatic elements would be produced in smaller quantities. To explore this idea, we use chemical evolution models generated using flexCE \citep{Andrews2017} to analyze how a galaxy's [Al/Mg]-[Mg/H], [Na/Mg]-[Mg/H], [Ca/Mg]-[Mg/H], and [Si/Mg]-[Mg/H] abundance patterns change when the model contains an upper mass cut-off of 30 solar masses such that these stars are not formed and do not contribute to the chemical evolution of the galaxy (Figure \ref{fig:almg}). We only changed this upper mass cut-off, otherwise using the parameters of their fiducial model, which used CCSNe yields from \citet{Limongi2006}, AGB yields from \citet{Karakas2010b}, and Type Ia SNe yields from the W70 model of \citet{Iwamoto1999}. We chose to analyze these 4 elements as they are among the best-fit elements to the MW. 

From the top panel of Figure \ref{fig:resid_mw_sum}, we would expect the mass cut-off model to show relative deficiencies in Al and Na, slight enhancement in Ca, and no change in Si. This is qualitatively consistent with what we see in Figure \ref{fig:almg} between the two models. [Ca/Mg] is enhanced only for the more chemically evolved stars, but recall that the MW 2-process model can only be applied to the most chemically evolved satellite galaxy stars as that is where the samples overlap. The models also suggest that we should see small residuals in [Si/Mg], which we do not observe. Obviously these models ultimately depend on the yields, which are often uncertain, but a top-light IMF might be something that is revisited once a dilution vector is taken into account and better understood. The DG 2-process model accurately predicting the elements for all satellite galaxies suggest that if the top-light IMF/mass-dependent outflows scenario were responsible for the MW 2-process model residuals, then all satellite galaxies would have to have the same IMF/outflows, which is perhaps unexpected given the range of stellar masses and environments probed by these satellite galaxies.

\subsection{Atypical SNe in the Satellite Galaxies}
\label{sec:sne}

Another explanation of why the MW 2-process model results in large satellite galaxy residuals for some elements and the DG 2-process model does not is that fundamentally different types of SNe are occurring between the two types of galaxies. These could be for both CC and Type Ia SNe, but given that all of the Fe-peak elements aside from Cr are under-predicted, perhaps the satellite galaxies are primarily polluted by a different type of Type Ia SNe. A range of literature works \citep[e.g.,][]{Kirby2019,Kobayashi2020b,delosReyes2022} find that the abundance patterns of some of the dwarf spheroidal galaxies are most consistent with being polluted by Sub-Chandrasekhar mass Type Ia SNe. Figure 10 of \citet{delosReyes2022} demonstrates that using Sub-Chandrasekhar mass Type Ia SNe yeilds from \citet{Leung2020} and \citet{Shen2018} with their best-fit chemical evolution model of the Sculptor dSph galaxy results in [Mn/Fe] and [Ni/Fe] abundance ratios that are $\sim$0.3-0.5 dex lower than with Chandrasekhar mass Type Ia SNe yields. However, it is not clear if Sub-Chandrasekhar mass Type Ia SNe are not occurring in the MW, with a recent study by \citet{Palla2021} demonstrating that a mixture of Sub-Chandrasekhar mass Type Ia SNe and Chandrasekhar mass Type Ia SNe match the MW abundance trends. Still, perhaps it is this mixture that is different between the MW and satellite galaxies, or perhaps even more generally, the mixture is different depending on the metallicity, i.e., more metal-poor Type Ia SNe are more likely to be Sub-Chandrasekhar mass. 

For all satellite galaxies studied here, the maximum $A_{\mathrm{Ia}}/A_{\mathrm{CC}}$ ratios (see Figure \ref{fig:2proc_ratio}) occur at slightly difference metallicities for each galaxy. Fornax and the SMC have a peak $A_{\mathrm{Ia}}/A_{\mathrm{CC}}$ at [Mg/H] $\sim$ -1.2, whereas Sgr has a peak $A_{\mathrm{Ia}}/A_{\mathrm{CC}}$ ratio at [Mg/H] $\sim$ -1.0 and a secondary peak at the most metal-rich stars. Moreover, the value of this ratio is much higher in Fornax than in Sgr, implying that at [Mg/H] = -1.2, Type Ia SNe dominated the chemical evolution of Fornax, and did so in a way that was unlike any of the other satellite galaxies. If different types of Type Ia SNe occurred at different metallicities, we might expect some of the Fe-peak elements to show residuals in the SMC, and especially Fornax, when using the DG 2-process model to predict their abundances. The bottom panel of Figure \ref{fig:resid_mw_sum} shows that, out of all the satellite galaxies, Fornax has the largest median residuals for Si, K, Ni, and Co, with the Ni residuals being the most significant given the small scatter relative to the magnitude of the meidan residual. Ni and Co have the largest residuals, and are elements produced in Type Ia SNe. However, Mn and Cr, elements that are also produced in large quantities by Type Ia SNe, do not show any residuals. Therefore, a metallicity-dependent Type Ia SNe explosion scenario would have to result in different Co and Ni, but not different Mn and Cr.

\section{Summary and Conclusions}
We have used two sets of stellar samples, low-Ia MW + high-Ia MW and low-Ia MW + high-Ia DG (Sgr), to derive two sets of abundance vectors for two separate 2-process models, which we refer to as the MW 2-process model and the DG 2-process model. If we use the MW 2-process model to predict the abundances of the more metal-rich ([Mg/H] $>$ -0.8) stars of the LMC, Sgr, and GSE, we find that the model under-predicts the abundance of Ce by $\sim$ 0.2 dex for the LMC and Sgr, and over-predicts the abundances of C+N, Na, Al, Ni, Mn, and Co for all 3 satellite galaxies. The MW 2-process model also slightly under-predicts the abundance of Ca for these 3 satellite galaxies by $\sim$ 0.1 dex. The MW 2-process model accurately predicts the abundances of O, Si, K, and Cr. There are also residuals for S and V, but these are elements that are less precisely measured by APOGEE, and we do not consider them in the discussion.

Despite the apparent failure of the MW 2-process model to predict many of the abundances of the satellite galaxies, if we use the abundances of Sgr to derive abundance vectors for a DG 2-process model, we find that the abundances for all of the satellite galaxies are reasonably well-predicted by this model. This means that, despite the wide variation in observed abundance patterns for each element between the galaxies as highlighted by \citet{Hasselquist2021}, each galaxy has the ``expected'' [X/Mg] abundance given its median [Fe/Mg] abundance at fixed [Mg/H]. Therefore, to the precision to which these abundances can be measured, the prompt and delayed processes that enriched the gas in the satellite galaxies occurred in a very similar way, but one or more of these processes are fundamentally different between the satellite galaxies in the MW.

Regardless of the interpretation in terms of enrichment processes, one can view the two-process model as a way of comparing the multi-element abundances of stellar populations that accounts for differences in overall metallicity and [$\alpha$/Fe] ratio.  Our main empirical finding (Figure \ref{fig:resid_mw_sum}) is that after accounting for these differences, the abundance patterns of the remaining 14 elements investigated here are similar among the five dwarf satellites but different between these satellites and the MW disk.

We have considered a number of possible explanations for this difference, and the full explanation may be a mix of these and other mechanisms.  The delayed contribution to Ce, and perhaps to other elements in our data set, is likely dominated by AGB stars rather than SNe Ia. The $+0.2$-dex residual for Ce in the LMC and Sgr implies that the ratio of AGB to Type Ia enrichment is higher in these systems than in the MW disk or GSE, for the common metallicity range.  This differing ratio could arise from different star formation histories, or perhaps from different relative retention of AGB and CC/Ia SNe products, or even from differences in binary star populations that lead to different numbers of SNe Ia per unit mass of star formation.  We find tentative evidence that Al enrichment is more closely tied to C+N rather than Mg, so that whatever mechanism leads to lower C+N in the satellites also leads to lower Al.

Many of the elements that show large residuals are those that also have metallicity-dependent process vectors, which in turn reflect metallicity-dependent yields.  Motivated by the two-infall scenario for MW chemical evolution \citep{Chiappini1997,Spitoni2019}, we hypothesize that a large inflow of pristine gas could have diluted the abundances of all elements in the ISM prior to the formation of the high-Ia, thin-disk population (Figure \ref{fig:dilution}).  In this case, the abundance {\it ratios} of low metallicity disk stars ([Mg/H]$\approx -0.6$) would be set by yields of a higher metallicity stellar population ([Mg/H]$\approx 0-0.2$), making them markedly different from those of DGs that reached [Mg/H]$\approx -0.6$ near the end of their evolution.  This scenario could be investigated more thoroughly in future studies by extending the 2-process model to allow for dilution and searching for signatures of this effect in disk populations of different age, metallicity, and location.

All of these explanations assume that the IMF-averaged yields of CCSNe and SNe Ia are the same between the DGs and the MW.  Another possibility is that these averaged yields are themselves different, even at the same metallicity, perhaps because of different IMF slopes or different ratios of sub-Chandrasekhar mass to Chandrasekhar-mass progenitors of SNe Ia.  This scenario requires a systematic difference between the MW and its satellites but close consistency among the satellites themselves --- only Fnx shows residuals relative as large as 0.1 dex for multiple elements (Ni, Co, Ce, and the less reliably measured Na) relative to the other DGs.  The impact of IMF changes could be studied with theoretical yield calculations to predict multi-element signatures of this effect, and differences in SNe Ia yields could be tested by focusing on elements whose production is most sensitive to progenitor properties.  If some of the chemical differences between the MW and its satellites can be conclusively tied to different yields, they will imply a remarkable connection between stellar scale astrophysics and galactic scale environment.

\begin{acknowledgments}
We thank the anonymous referee for their comments and suggestions that greatly improved this manuscript. SH is supported by STScI DDRF grant D0001.82510. This research made use of Astropy \footnote{http://www.astropy.org} a community-developed core Python package for Astronomy \citep{astropy:2013,astropy:2018}, SciPy \citep*{SciPy}, NumPy \citep{NumPy}, and Matplotlib \citep{Hunter:2007}.

\end{acknowledgments}

\bibliography{ref_og}{}
\bibliographystyle{aasjournal}

\appendix

\section{Results from the $\log{\lowercase{g}}$-matched MW sample} \label{sec:appendix}

\begin{figure*}
    \includegraphics[width=\textwidth]{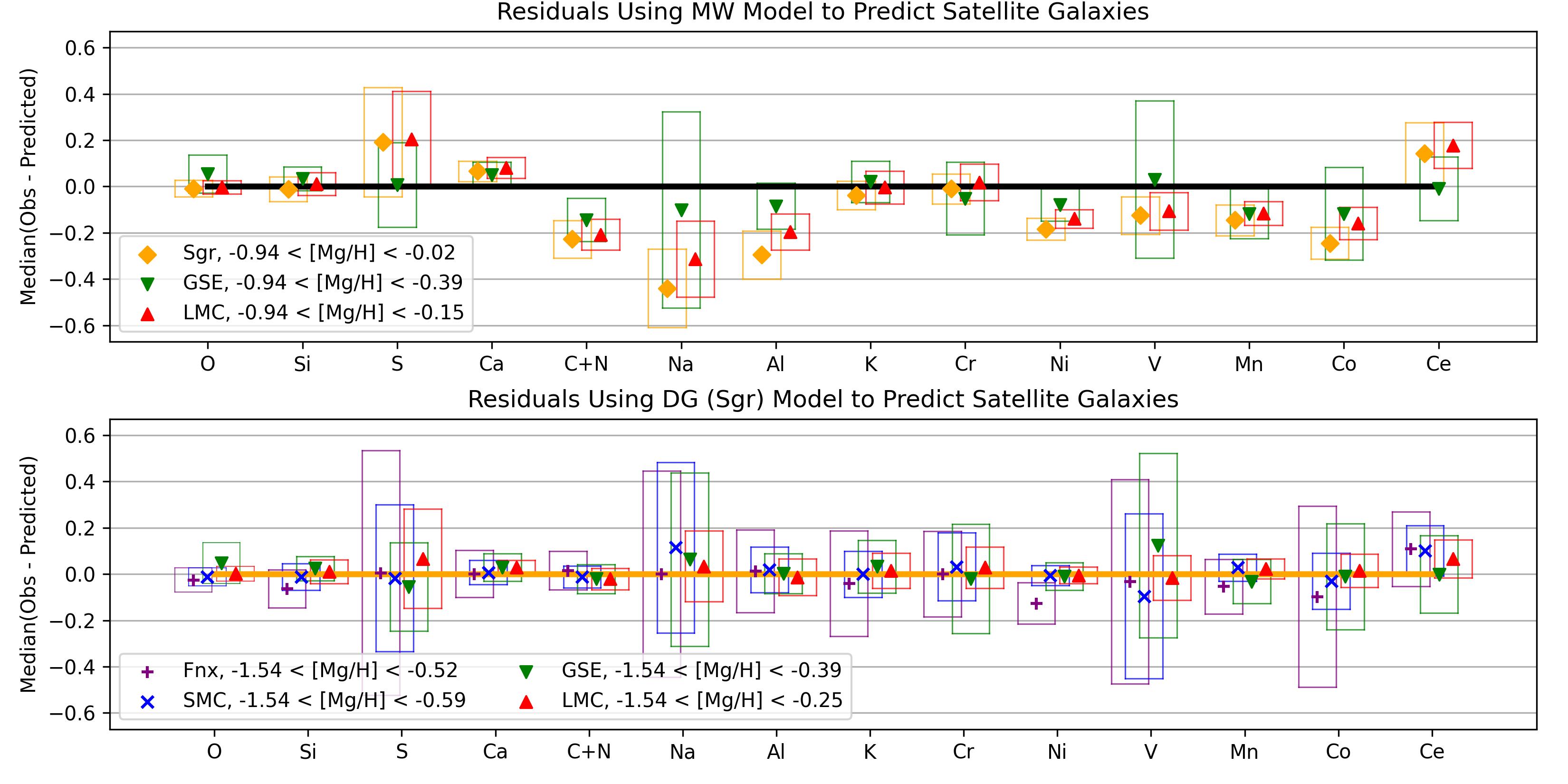}    
    \caption{Same as Figure \ref{fig:resid_mw_sum} but for the $\log{g}$-matched MW sample instead of the T$_{\rm eff}$-matched MW sample. \label{fig:resid_mw_sum_logg}}
\end{figure*}

In Section \ref{sec:data} we discuss using a T$_{\rm eff}$-matched MW sample as well as a $\log{g}$-matched sample. Figure \ref{fig:param_select} shows that when we match on one of these parameters, the distribution of the other parameter does differ from the distribution of the satellite galaxies. We have primarily focused the results and discussion on the T$_{\rm eff}$-matched MW sample, but we show in Figure \ref{fig:resid_mw_sum_logg} that the residual abundance patterns we get from deriving 2-process model parameters from the $\log{g}$-matched sample are nearly identical to the results from the T$_{\rm eff}$-matched MW sample.

\end{document}